\documentclass[fleqn,usenatbib]{mnras}

\usepackage{newtxtext,newtxmath}

\newcommand{\Msun}{\mbox{$\mathrm{M}_{\odot}$}}

\usepackage[T1]{fontenc}

\DeclareRobustCommand{\VAN}[3]{#2}
\let\VANthebibliography\thebibliography
\def\thebibliography{\DeclareRobustCommand{\VAN}[3]{##3}\VANthebibliography}

\usepackage{graphicx}	
\usepackage{amsmath}	

\title[DAHe white dwarfs from the DESI Survey]{DAHe white dwarfs from the DESI Survey}

\author[C.\,J.\,Manser et al.]{\noindent
Christopher J. Manser,$^{1,2}$\thanks{E-mail: c.j.manser92@googlemail.com}
Boris T. G\"ansicke,$^{2}$
Keith Inight,$^{2}$
Akshay Robert,${^{1,3}}$
S.~Ahlen,${^{4}}$
\newauthor
C.~Allende~Prieto,${^{5,6}}$
D.~Brooks,${^{3}}$
A.P.~Cooper,${^{7}}$
A.~de la Macorra,${^{8}}$
A.~Font-Ribera,${^{9}}$
K.~Honscheid,${^{10,11}}$
\newauthor
T.~Kisner,${^{12}}$
M.~Landriau,${^{12}}$
Aaron M. Meisner,${^{13}}$
R.~Miquel,${^{9,14}}$
Jundan Nie,${^{15}}$
C.~Poppett,${^{12,16,17}}$
\newauthor
Gregory~Tarl\'{e},${^{18}}$
Zhimin~Zhou${^{15}}$
\\
$^{1}$ Astrophysics Group, Department of Physics, Imperial College London, Prince Consort Rd, London, SW7 2AZ, UK \\
$^{2}$ Department of Physics, University of Warwick, Coventry CV4 7AL, UK \\
$^{3}$ Department of Physics and Astronomy, University College London, London, WC1E 6BT, UK \\
$^{4}$ Physics Dept., Boston University, 590 Commonwealth Avenue, Boston, MA 02215, USA \\
$^{5}$ Instituto de Astrof\'{i}sica de Canarias, C/ Vía L\'{a}ctea, s/n, E-38205 La Laguna, Tenerife, Spain \\  
$^{6}$ Universidad de La Laguna, Dept. de Astrof\'{\i}sica, E-38206 La Laguna, Tenerife, Spain \\
$^{7}$ Institute of Astronomy and Department of Physics, National Tsing Hua University, 101 Kuang-Fu Rd. Sec. 2, Hsinchu 30013, Taiwan \\
$^{8}$ Instituto de F\'{\i}sica, Universidad Nacional Aut\'{o}noma de M\'{e}xico,  Cd. de M\'{e}xico  C.P. 04510,  M\'{e}xico \\
$^{9}$ Institut de F\'{i}sica d’Altes Energies (IFAE), The Barcelona Institute of Science and Technology, Campus UAB, 08193 Bellaterra Barcelona, Spain \\
$^{10}$ Center for Cosmology and AstroParticle Physics, The Ohio State University, 191 West Woodruff Avenue, Columbus, OH 43210, USA \\
$^{11}$ Department of Physics, The Ohio State University, 191 West Woodruff Avenue, Columbus, OH 43210, USA \\
$^{12}$ Lawrence Berkeley National Laboratory, 1 Cyclotron Road, Berkeley, CA 94720, USA \\
$^{13}$ NSF's NOIRLab, 950 N. Cherry Ave., Tucson, AZ 85719, USA \\
$^{14}$ Instituci\'{o} Catalana de Recerca i Estudis Avan\c{c}ats, Passeig de Llu\'{\i}s Companys, 23, 08010 Barcelona, Spain \\
$^{15}$ National Astronomical Observatories, Chinese Academy of Sciences, A20 Datun Rd., Chaoyang District, Beijing, 100012, P.R. China \\
$^{16}$ Space Sciences Laboratory, University of California, Berkeley, 7 Gauss Way, Berkeley, CA  94720, USA \\
$^{17}$ University of California, Berkeley, 110 Sproul Hall \#5800 Berkeley, CA 94720, USA \\
$^{18}$ University of Michigan, Ann Arbor, MI 48109, USA\\
}
\date{Accepted 2023 March 7. Received 2023 March 6; in original form 2023 February 2}

\pubyear{2023}

\begin{document}
\label{firstpage}
\pagerange{\pageref{firstpage}--\pageref{lastpage}}
\maketitle

\begin{abstract}
A new class of white dwarfs, dubbed DAHe, that present Zeeman-split Balmer lines in emission has recently emerged. However, the physical origin of these emission lines remains unclear. We present here a sample of 21 newly identified DAHe systems and determine magnetic field strengths and (for a subset) periods which span the ranges of $\simeq$\,6.5\,--\,147\,MG and $\simeq$\,0.4\,--\,36\,h respectively. All but four of these systems were identified from the Dark Energy Spectroscopic Instrument (DESI) survey sample of more than 47\,000 white dwarf candidates observed during its first year of observations. We present detailed analysis of the new DAHe WD\,J161634.36+541011.51 with a spin period of 95.3\,min, which exhibits an anti-correlation between broadband flux and Balmer line strength that is typically observed for this class of systems. All DAHe systems cluster closely on the \textit{Gaia} Hertzsprung-Russell diagram where they represent $\simeq$\,1\,per\,cent of white dwarfs within that region. This grouping further solidifies their unexplained emergence at relatively late cooling times and we discuss this in context of current formation theories. Nine of the new DAHe systems are identifiable from SDSS spectra of white dwarfs that had been previously classified as featureless DC-type systems. We suggest high $S/N$, unbiased observations of DCs as a possible route for discovering additional DAHe systems. 
\end{abstract}



\section{Introduction}

The study of magnetic fields and its effect on matter is important across scientific disciplines and their applications, such as magnetic resonance imaging in the field of medicine \citep{lauterbur73-1}. The highest continuous field strength generated in the laboratory is recorded at 0.455\,MG \citep{hahnetal19-1}, but to investigate fields beyond this we must study and characterise astronomical sources. White dwarfs, the remnant cores left over from the stellar evolution of main sequence stars up to masses of $\simeq$\,8\,--\,10\,\Msun, have weaker fields than the more massive neutron stars \citep{truemperetal78-1, olausen+kaspi14-1}, but are significantly easier to observe and characterise (for reviews, see \citealt{wickramasinghe+ferrario00-1, ferrarioetal20-1}).

Since the discovery of the first magnetic white dwarf, Grw+70$^{\rm o}$8247, more than half a century ago \citep{kempetal70-1}, several hundred systems are now known with fields in the range of $\sim$\,10$^{-3}$\,--\,10$^{3}$\,MG \citep{kuelebietal09-1, ferrarioetal15-1,bagnulo+landstreet22-1}. These white dwarfs have been predominantly identified by the Zeeman-splitting of spectroscopic absorption features \citep{preston70-1,angeletal74-2,schmidtetal03-1}, although other methods can be used such as spectropolarimetric measurements which can detect the weakest kilogauss fields in bright nearby white dwarfs with broadband circular polarisation \citep{kempetal70-1,bagnulo+landstreet20-1, berdyuginetal22-1,berdyuginetal23-1} or with polarisation in spectral lines \citep{angel+landstreet70-1,cuadradoetal04-1, jordanetal07-1,landstreet+bagnulo19-1}, and the detection of cyclotron emission/absorption from accreting systems with a magnetic white dwarf primary \citep{visvanathan+wickramasinghe81-1,schwopeetal90-1,baileyetal91-1, campbelletal08-1}.

The generation of magnetic fields in isolated white dwarfs cannot come from a single mechanism \citep{bagnulo+landstreet21-1, bagnulo+landstreet22-1}, and multiple channels have been proposed which include: (i) Fossil fields strengthened by magnetic flux conservation as stars evolve into white dwarfs \citep{woltjer64-1,landstreet67,toutetal04-1}, (ii) convective dynamos driven in binary mergers \citep{reghos+tout95-1, toutetal08-1, nordhausetal11-1, garcia-berroetal12-1, wickramasingheetal14-1}, or by giant-planet engulfment \citep{siess+livio99-1, farihietal11-2}, and (iii) convective dynamos driven by crystallization of the cores of white dwarfs as they cool down \citep{vanhorn68-1, isernetal17-1, schreiberetal21-1, schreiberetal21-2, schreiberetal22-1, ginzburgetal22-1}. 

A rare subset of magnetic white dwarfs show Zeeman-split Balmer \textit{emission} (DAHe), first identified in the white dwarf GD\,356 \citep{greenstein+mccarthy85-1}. Three DAHe white dwarfs are currently known\footnote{Two potentially additional systems show Balmer-line emission but no detectable magnetic field; WD\,J0412+7549 \citep{tremblayetal20-1, waltersetal21-1} and WD\,J1653--1001 \citep{obrienetal23-1}, dubbed DAe white dwarfs.} \citep{redingetal20-1,gansickeetal20-1,waltersetal21-1}, which all appear to be near the age at which crystallization sets in \citep{schreiberetal21-2} potentially explaining both their emergence in the white dwarf cooling sequence and their magnetic fields. Additionally, these DAHe systems are fast rotators with spin periods in the range of 0.09\,--\,15.3\,h. Isolated white dwarfs typically have spin periods of 1\,--\,3\,d \citep{hermesetal17-1}, and the accretion of planetary material by these white dwarfs could provide a source of angular momentum to spin them up \citep{stephanetal21-1}. The presence of planetary material around white dwarfs is not unexpected, as the survival of planetary systems into the white dwarf phase of their host star is evidenced through the presence, destruction, and accretion of planetary bodies \citep{zuckerman+becklin87-1, zuckermanetal03-1, jura03-1, gaensickeetal06-3, vanderburgetal15-1, gaensickeetal19-1, vanderburgetal20-1, blackmanetal21-1, cunninghametal22-1, gaia22-1}. 

While a crystallization-driven convective dynamo could produce a magnetic field in these white dwarfs, the generation mechanism of Zeeman-split \textit{emission} is still uncertain. Observations show that the Zeeman-split emission is strongest at photometric minimum in these DAHe systems. This coincidence suggests that these white dwarfs host a temperature-inverted, optically thin emission region sitting above a photospheric dark region. The photometric variability induced by these dark regions has been observed in many magnetic white dwarfs \citep{wickramasinghe+ferrario00-1}, and are thought to be generated by the change in the inhomogeneous field distribution over the visible surface of the white dwarf while it rotates. The resulting photometric variability has previously been attributed to a reduction in the temperature in areas of high magnetic field strength, akin to a Sun-spot. However, recent theoretical works suggest that photometric variability in magnetic radiative main-sequence stars is due to changes in their emergent spectrum rather than their bolometric flux \citep{fuller+mathis23-1}, although further work is needed to extend this result to white dwarfs. 

For GD\,356, the size of both of these emitting and photometrically darker regions cover $\simeq$\,10\,per\,cent of the white dwarf surface \citep{ferrarioetal97-1, brinkworthetal04-1}. One potential explanation for these systems is the unipolar inductor model \citep{goldreich+lyndenbell69-1, lietal98-1, wickramasingheetal10-1}, whereby a close-in conductive body, such as a planetary core, induces a current that heats up the white dwarf atmosphere at the magnetic poles. However, this model is debated \citep{waltersetal21-1}, and other possibilities could include the presence of an intrinsic chromosphere in these white dwarfs. As only three confirmed DAHe systems are currently known; enlarging this class of stars is crucial for understanding these white dwarfs and determining their origin.

In this paper, we report the identification of 21 new DAHe systems. Seventeen of these systems were identified by the Dark Energy Spectroscopic Instrument (DESI; \citealt{DESI16-1, DESI16-2}) on the Mayall 4\,m telescope at Kitt Peak National Observatory (KPNO), and four were discovered from re-inspection of archival Sloan Digital Sky Survey (SDSS; \citealt{yorketal00-1, eisensteinetal11-1,blantonetal17-1}) spectroscopy of featureless DC white dwarfs. In Section\,\ref{s2} we present detailed observations for the DAHe, WD\,J161634.36+541011.51 (hereafter WD\,J1616+5410), which shows Zeeman-emission and photometric variability on a $\simeq95.3$\,min period. In Section\,\ref{s3} we present our results of analysing the presence of a magnetic field and photometric variability in WD\,J1616+5410. In Section\,\ref{s4} we give details on the 21 DAHe white dwarfs we have identified. In Section\,\ref{s5} we calculate the occurrence rates of DAHe white dwarfs and discuss the potential theories for why these systems cluster at late cooling times on the \textit{Gaia} Hertzsprung-Russell diagram (HRD). Finally we summarise our findings in Section\,\ref{s6}.

\section{Observations}\label{s2}

\begin{figure*}
\centerline{\includegraphics[width=2\columnwidth]{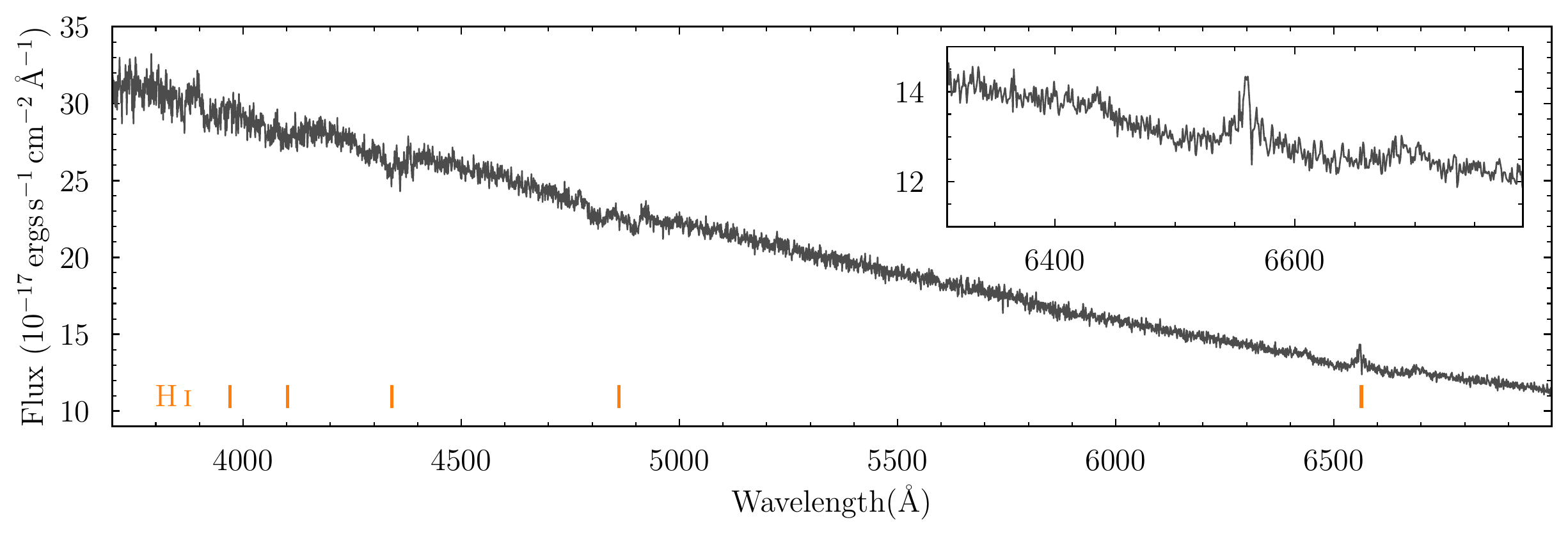}}
\caption{\label{f-WDJ1616+5410} The coadded spectrum of WD\,J1616+5410 in gray with the rest wavelengths of the Balmer series denoted by orange tabs. An inset highlights the H$\alpha$ region where Zeeman-split emission is seen. The spectrum is featureless beyond 7\,000\,\AA.}
\end{figure*}

\begin{table}
\centering
\caption{Log of DESI spectroscopy for WD\,J1616+5410, where HJD is given for the time the spectrograph shutter opens.  \label{t-spectral_dates}}
\begin{tabular}{llllll}
\hline
\# & HJD [d]         & Exp.     & Night      & Exp.  & Fibre \\
   &                 & time [s] &            & ID    &       \\
\hline                              
1  & 2459288.9659351 &  900.1  & 2021 03 14 & 80503 & 4903  \\ 
2  & 2459288.9787686 &  900.1  & 2021 03 14 & 80504 & 4903  \\ 
3  & 2459288.9909529 &  900.1  & 2021 03 14 & 80505 & 4903  \\ 
4  & 2459289.0030943 &  900.1  & 2021 03 14 & 80506 & 4903  \\ 
5  & 2459289.0152451 &  900.1  & 2021 03 14 & 80507 & 4903  \\ 
6  & 2459311.8965222 &  636.4  & 2021 04 06 & 83738 & 4631  \\ 
7  & 2459312.9428680 &  259.5  & 2021 04 07 & 83886 & 4631  \\ 
8  & 2459323.9773956 &  913.5  & 2021 04 18 & 85357 & 4839  \\ 
9  & 2459323.9888389 &  956.2  & 2021 04 18 & 85358 & 4839  \\ 
10 & 2459324.9054519 &  326.7  & 2021 04 19 & 85511 & 4839  \\ 
11 & 2459337.9370142 &  1431.0 & 2021 05 02 & 86988 & 4921  \\
12 & 2459338.9186979 &  800.9  & 2021 05 03 & 87122 & 4921  \\ 
13 & 2459340.9281868 &  835.3  & 2021 05 05 & 87379 & 4530  \\ 
\hline
\end{tabular}
\end{table}

\begin{table*}
\centering
\caption{Log of ZTF and LT photometry for WD\,J1616+5410. \label{t-photo_log}}
\begin{tabular}{lrrrrrr}
\hline
Telescope & Night & Band & First exp. [HJD] & last exp. [HJD] & Exp. time [s] & \# of exp. \\
\hline                            
ZTF & Multiple    & $g$ & 2458203.895314 & 2459641.895052 & 30 & 1124 \\
ZTF & Multiple    & $r$ & 2458198.893930 & 2459641.868801 & 30 & 1202 \\
LT  & 2022 May 26 & $g$ & 2459726.541376 & 2459726.704541 & 60 & 180  \\ 
LT  & 2022 May 28 & $g$ & 2459728.509526 & 2459728.563321 & 60 & 60   \\ 
LT  & 2022 May 31 & $g$ & 2459731.487386 & 2459731.650531 & 60 & 179  \\ 
\hline
\end{tabular}
\end{table*}

\begin{figure*}
\centerline{\includegraphics[width=2\columnwidth]{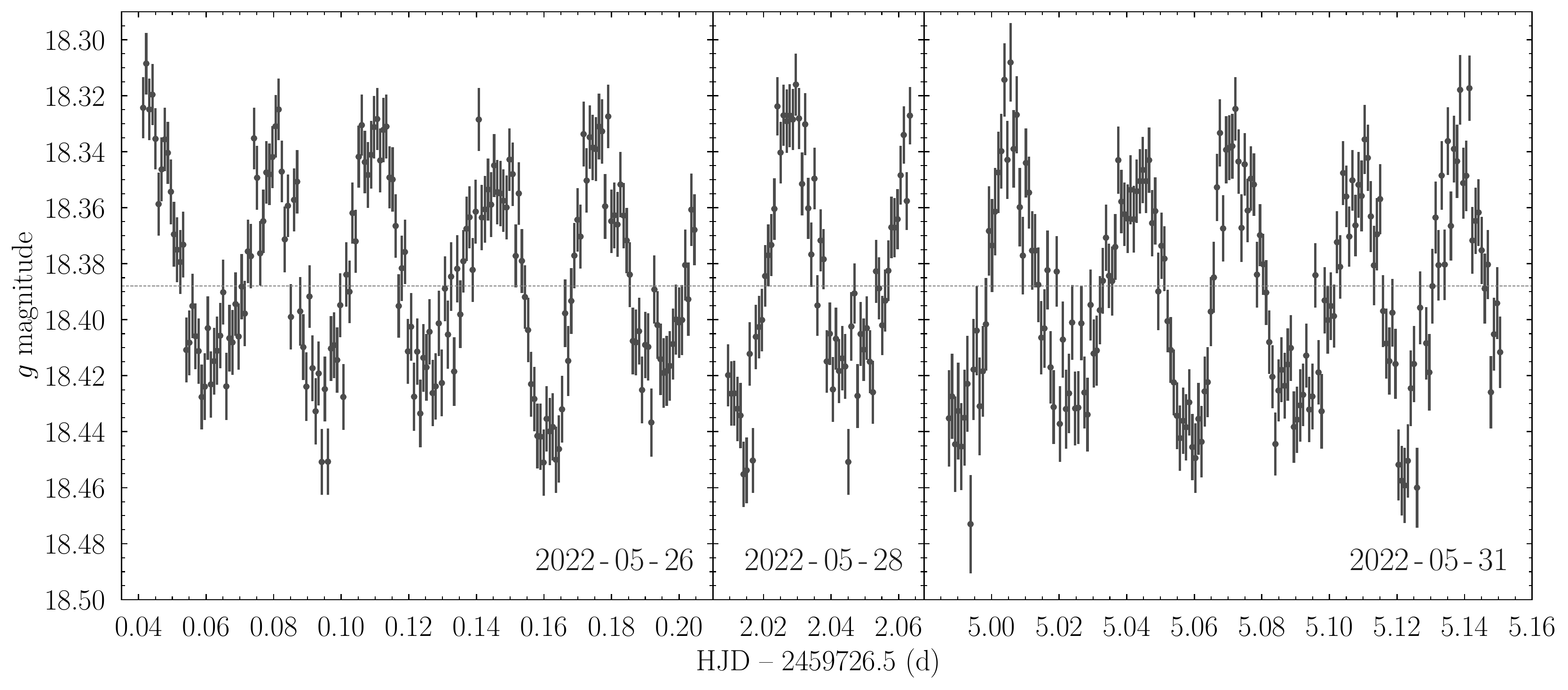}}
\caption{\label{f-lc-LT} LT photometry of WD\,J1616+5410. The horizontal dashed line corresponds to the median value of the LT photometry.}
\end{figure*}  

\begin{figure}
\centerline{\includegraphics[width=1\columnwidth]{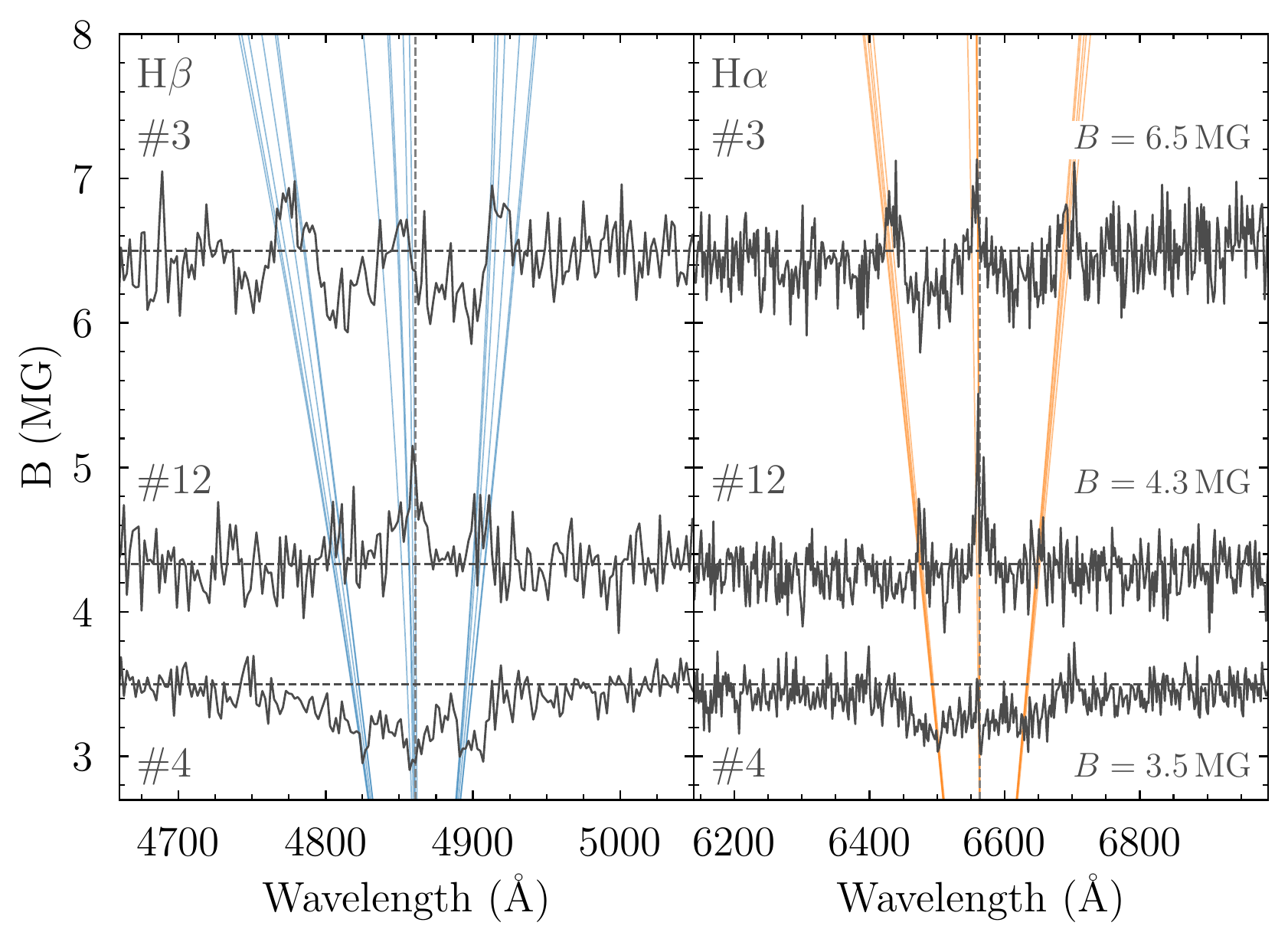}}
\caption{\label{f-3_4} Continuum normalised spectra of WD\,J1616+5410 showing the H$\beta$ and H$\alpha$ regions of spectra \#3, \#4, and \#12 (see Table\,\ref{t-spectral_dates}), where the continuum flux has been multiplied by the estimated field strength for each spectrum. The emission spectra \#3 and \#12 are observed in the two photometric minima and originate from the two magnetic poles of the white dwarf (see Fig.\ref{f-toymodel}), whereas spectrum \#4 is observed during a photospheric maximum where the poles are most out of view (see Fig.\,\ref{f-WDJ1616_phasefold_g}).}
\end{figure} 

\begin{figure*}
\centerline{\includegraphics[width=2\columnwidth]{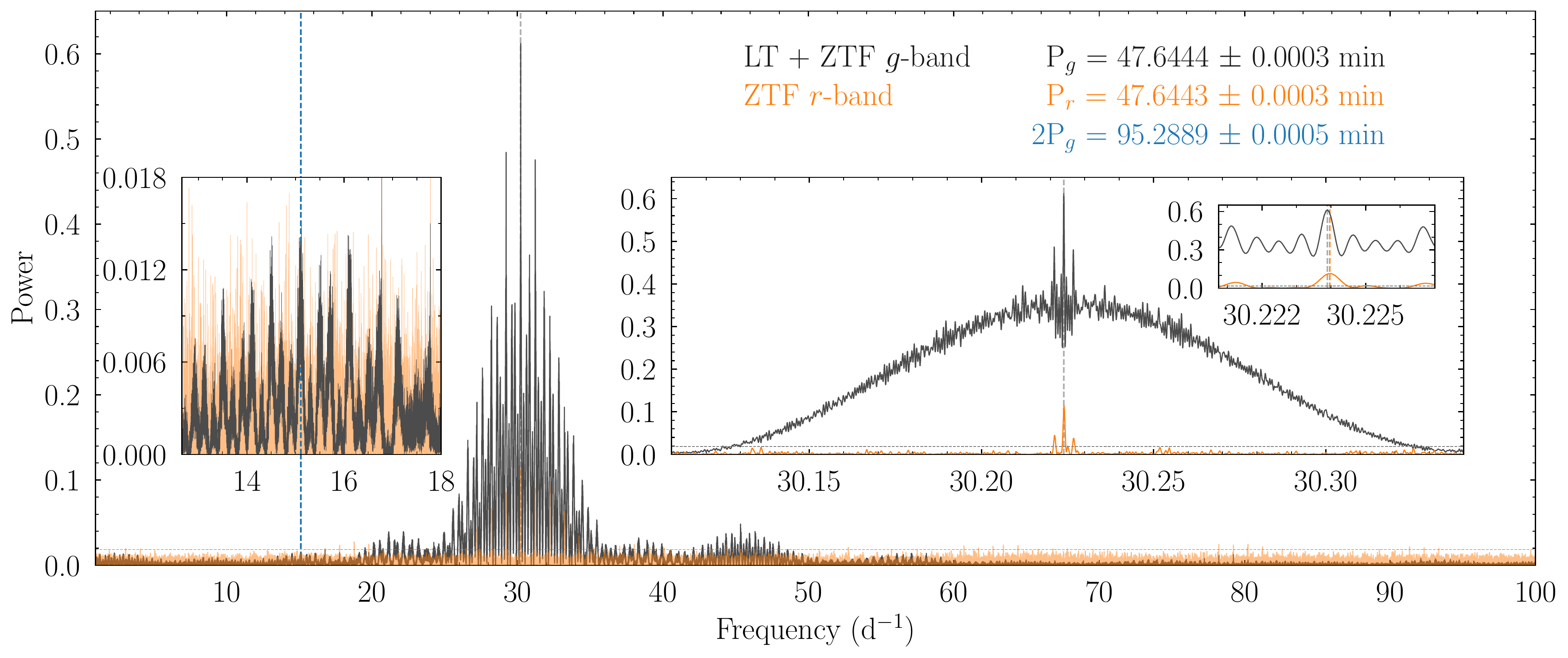}}
\caption{\label{f-combined_periodogram} Lomb-Scargle periodogram calculated from combined ZTF and LT $g$-band photometry (gray) and ZTF $r$-band photometry (orange). Insets show zoom-ins on the strongest signal, with a period of $P_g$\,=\,47.6444\,$\pm$\,0.0003\,min and twice that period at $2P_g$\,=\,95.2889\,$\pm$\,0.0005\,min which appears to be the true period from the analysis of the magnetic field (see Figs.\,\ref{f-3_4}\,\&\,\ref{f-WDJ1616_phasefold_g}). Vertical dashed lines are placed at the frequencies corresponding to the noted periods, and the horizontal dashed line corresponds to a false alarm probability of 1\,per\,cent for signals obtained from the combined ZTF and LT $g$ band photometry.}
\end{figure*} 

\subsection{Spectroscopy}

\subsubsection{DESI}

DESI on the Mayall 4\,m telescope at Kitt Peak National Observatory (KPNO) is a multi-object spectroscopic instrument capable of collecting fiber spectroscopy on up to $\simeq5000$ targets per pointing \citep{abareshietal22-1}. The fibers are positioned by robot actuators and are grouped into ten petals which feed ten identical three-arm spectrographs, each spanning 3600\,--\,9824\,\AA\ at a FWHM resolution of $\simeq$\,1.8\,\AA. The inter-exposure sequence which includes telescope slewing, spectrograph readout and focal plane reconfiguration can be completed in as little as $\simeq$\,2\,min \citep{abareshietal22-1}. DESI started main survey operations on 2021 May 14 and will obtain spectroscopy of more than 40\,million galaxies \citep{zhouetal20-1, zhouetal22-1, raichooretal20-1, raichooretal22-1} and quasars \citep{yecheetal20-1, chaussidonetal22-1} over five years to explore the nature of dark matter. During sub-optimal observing conditions (for example, poor seeing or high lunar illumination), observations switch focus to nearby bright galaxies \citep{ruiz-maciasetal20-1,hahnetal22-1} and stars \citep{allendeprietoetal20-1, cooperetal22-1}. The DESI Early Data Release (EDR, DESI collaboration et al. 2022, in preparation) contains approximately $\simeq$\,4\,400 white dwarf candidates \citep{cooperetal22-1} from the \cite{fusilloetal19-1} \textit{Gaia} DR2 sample, and a full description of the reduction pipeline is given by \cite{guyetal22-1}. The first year of DESI observations will constitute the DESI Data Release 1 (DR1), and combined with the DESI EDR contain over 47\,000 white dwarf candidate. The EDR will contain survey validation data (DESI collaboration et al. 2023, in preperation), where targets often have more repeat observations than in main survey operations.

We identified a total of 17 DAHe systems through visual inspection of the DESI EDR and DR1 samples (see Section\,\ref{s4}, which includes the white dwarf WD\,J1616+5410 (Fig.\,\ref{f-WDJ1616+5410}). As WD\,J1616+5410 was observed during survey validation many repeat spectra were taken in varied conditions (leading to a visible spread in the signal-to-noise ratios, $S/N$), resulting in thirteen individual exposures collected by DESI from 2021 March 14 to 2021 May 5, which are presented in Table\,\ref{t-spectral_dates}. The first five exposures were taken consecutively with an exposure time of $\simeq$\,900\,s, and show clear variation of the Balmer features, exhibiting both emission and absorption features (see Fig.\,\ref{f-all_spec}). Given the variations detected in the DESI spectroscopy, we inspected archival Zwicky Transient Facility (ZTF) data to probe for photometric variability.

\subsubsection{SDSS}\label{s2SDSS}

The SDSS has been taking multi-band photometry and multi-fibre spectroscopy since 2000, using a 2.5\,m telescope located at the Apache Point Observatory in New Mexico \citep{gunnetal06-1}. We retrieved the archival SDSS spectroscopy (DR17, \citealt{abdurroufetal22-1}) of 2621 white dwarfs classified as ``DC'' within the \textit{Gaia}/SDSS white dwarf catalogue of  \cite{fusilloetal19-1}, using  the \textsc{spectral\_class} identifier. We identified four new DAHe systems from this sample which are discussed further in Section\,\ref{s4SDSS}.

\subsection{Time-series photometry}

\subsubsection{ZTF}

ZTF is a robotic time-domain survey using the Palomar 48-inch Schmidt Telescope \citep{bellmetal19-1, mascietal19-1}. Utilising a 47\,$\mathrm{deg}^2$ field of view, ZTF can scan the entire sky in $\simeq$\,two days, making it a powerful survey for identifying photometrically variable sources at optical wavelengths. The ZTF observed WD\,J1616+5410 and we collected the photometry provided by Data Release 11 \citep{mascietal19-1}. Between 2018 March to 2022 March, ZTF obtained a total of 1124 and 1202 30\,s exposures in the $g$-band and $r$-band, respectively (Table\,\ref{t-photo_log}). We computed a preliminary Lomb-Scargle periodogram \citep{lomb76-1, scargle82-1} using the python package \texttt{astropy.timeseries} \citep{astropy-1,astropy-2,astropy-3}, and identified a strong periodic signal at $\simeq$\,47.6\,min, which spurred us on to collect additional photometry on the Liverpool Telescope (LT) to confirm the variability.

\subsubsection{LT}

The LT is a 2\,m robotic telescope situated on the island of La Palma \citep{steeleetal04-1}. We obtained three nights of data on the LT, on 2022 May 26, May 28, and May 31, collecting 419 60\,s exposures totaling $\simeq$\,7\,h (Table\,\ref{t-photo_log}). We used the IO:O imager with the SDSS-$g$ filter providing variability information over the wavelength range 4000\,--\,5500\,\AA, covering H$\beta$, H$\gamma$, and H$\delta$ where photometric variability is expected for magnetic white dwarfs. We used the standard LT pipeline to provide bias subtraction and flat fielding. Differential photometry based on Sextractor \citep{bertin+arnouts96-1} was then used to extract the light curve by comparing the flux with the comparison star, SDSS\,J161657.97+540954.4. The LT light-curve of WD\,J1616+5410 exhibits clear periodic behaviour (Fig.\ref{f-lc-LT}), with variations over a magnitude range of $\simeq$\,0.14\,mag on a period in agreement with that identified from the ZTF photometry, although there are clear changes in the strength at maxima and minima.

\section{Results}\label{s3}

\subsection{Magnetic field variations}

Figure\,\ref{f-all_spec} reveals clear variations in the H$\beta$ and H$\alpha$ features where spectra show both three-component absorption (e.g. \#2, \#4, \#6) and emission (e.g. \#3, \#12, \#13), a behaviour that has also been reported for the DAHe white dwarf WD\,J125230.93--023417.72 (hereafter WD\,J1252--0234; \citealt{redingetal20-1}). This is due to Zeeman-splitting of the energy levels for H$\beta$ and H$\alpha$ in the presence of a magnetic field. In spectrum \#4, the absorption features are consistent with the linear Zeeman-splitting regime where the three $\sigma^{-}$, $\pi$, and $\sigma^{+}$ components are clearly seen. As the magnetic field strength increases, the energy degeneracy due to orbital angular momentum, $l$, will eventually be lifted and these three components will split further, resulting in 18 (15) transitions in H$\beta$ (H$\alpha$), where the 2p$_{-1} \rightarrow 4\textrm{d}_{-1}$ and 2p$_{1} \rightarrow 4\textrm{d}_{1}$ (2p$_{-1} \rightarrow 3\textrm{d}_{-1}$ and 2p$_{1} \rightarrow 3\textrm{d}_{1}$) transitions are still degenerate \citep{henry+oconnell85-1}. This is known as the quadratic Zeeman effect and is apparent in the H$\beta$ region of spectrum \#3, where the central $\pi$ component is blue-shifted, in addition to all three emission features having broader profiles consistent with the lifting of energy degeneracy. 

We measured the magnetic field strength, $B$, for spectra \#3, \#4, and \#12, as they were the clearest profiles recovered from the DESI spectroscopy, and assume $B$ is constant throughout the emitting or absorbing area of the white dwarf atmosphere. We used the transition wavelengths as a function of $B$ provided by \cite{schimeczek+wunner14-1, schimeczek+wunner14-2} and compared these to the Zeeman-split profiles (Fig.\,\ref{f-3_4}), where we obtained $B$ values of 6.5\,$\pm$\,0.1\,MG, 4.3\,$\pm$\,0.1\,MG, and 3.5\,$\pm$\,0.2\,MG for spectra \#3, \#4, and \#12 respectively. Spectra \#3 and \#4 were taken sequentially with only $\simeq$\,3.5\,min of downtime (consistent with the time taken for the inter-exposure sequence), and suggest that the white dwarf is rotating fairly rapidly and that a localised emitting hot-spot is rotating in and out of view. The hotspot must be sufficiently localised as the width of the $\sigma$ profiles do not allow for a significant variation in $B$, which has been previously seen for the DAHe WD\,J121929.50+471522.94 (hereafter WD\,J1219+4715; \citealt{gansickeetal20-1}). For a centred-dipolar field configuration on the white dwarf, the field strength near the magnetic equator is a factor two lower than the field strength at the poles \citep{achilleosetal92-2}, which is in rough agreement with the variations between spectra \#3 and \#4. Spectrum \#12 was taken $\simeq$\,50\,d after \#3 and \#4, but is suggestive of two seperate localised emitting hot-spots with two different field strengths i.e an offset dipole or quadrupolar field \citep{martin+wickramasinghe84-1, achilleos+wickramasinghe89-1, achilleosetal92-2}. If this interpretation is correct and the dipole is simply offset in the direction of the magnetic axis \citep{achilleosetal92-2}, we expect the variation between spectrum \#3 and spectrum \#12 to be periodic and separated by $\simeq$\,0.5 in phase.

\begin{figure*}
\centerline{\includegraphics[width=2\columnwidth]{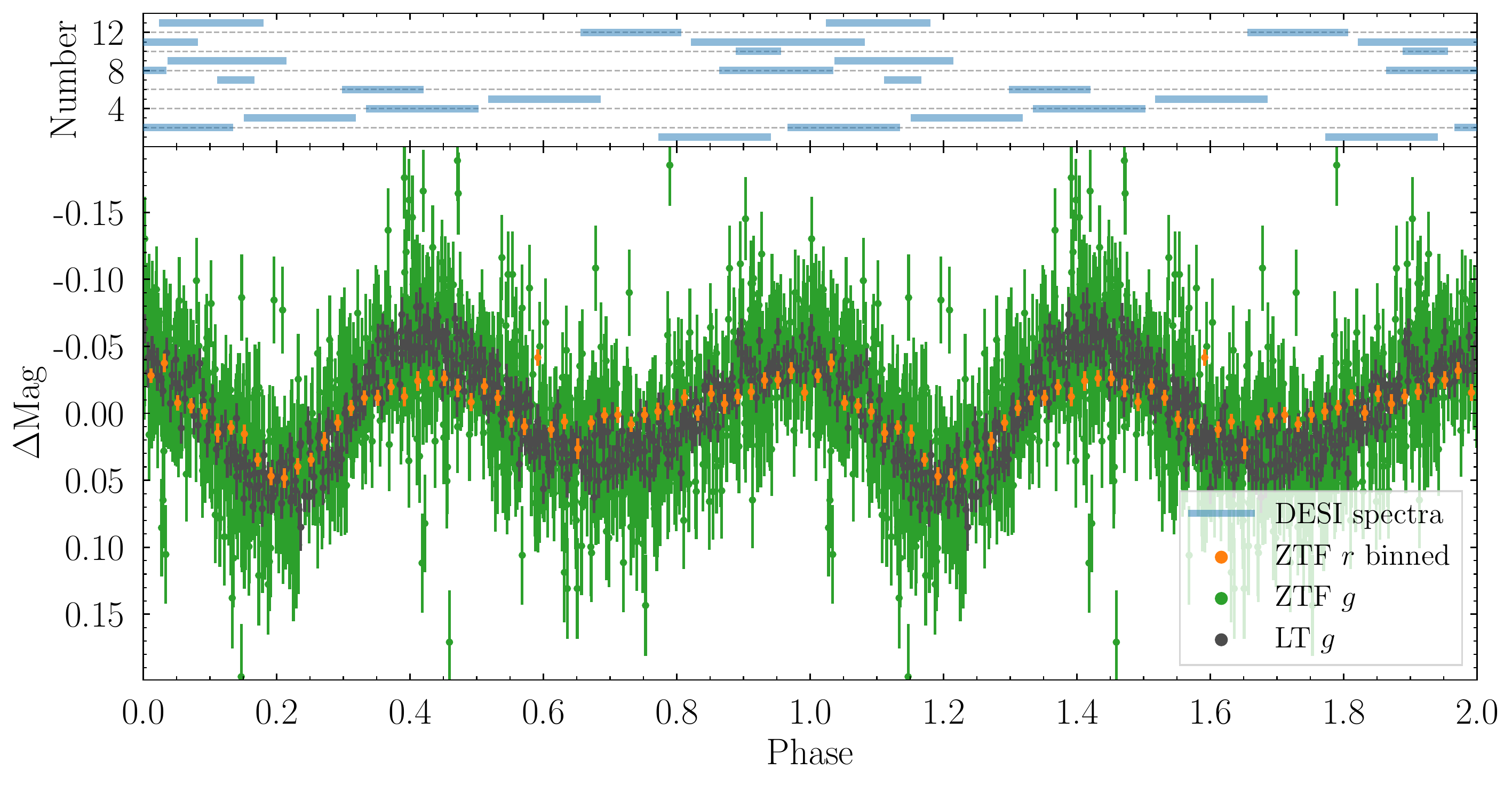}}
\caption{\label{f-WDJ1616_phasefold_g} Bottom: LT and ZTF $g$ band light curves folded onto a period of 95.288828\,$\pm$\,0.000023\,min, twice the period identified from Fig.\,\ref{f-combined_periodogram}. We also include ZTF $r$ band data phase-folded onto the same period and binned into 0.02 phase bins. Top: Horizontal blue tabs show the span of the DESI spectra of WD\,J1616+5410, where number of each tab matches the ``\#" column of Table.\,\ref{t-spectral_dates}. Two phase cycles are shown for clarity.}
\end{figure*}

\subsection{Period analysis}

With the combined set of ZTF and LT data, we computed Lomb-Scargle periodograms for the ZTF $r$-band data and the LT and ZTF $g$-band data. Owing to an offset between the ZTF and LT $g$-band data of about 0.05\,mag, we subtracted the median magnitude of each data-set separately before combining them. The resulting periodograms are shown in Fig.\,\ref{f-combined_periodogram}, where the 47.6444\,$\pm$\,0.0003\,min signal is immediately recovered from both the $g$-band and $r$-band data with excellent agreement between the two. We determined the error on the period by calculating the standard deviation of a Gaussian profile fitted to the signal peak in the periodogram. There are potential alias periods present in the ZTF and LT data, with some at a frequency half that of the dominant signal. Given the variation in maxima and minima seen in the LT light-curve, we inspected phase-folded light-curves on both a 47.6444\,$\pm$\,0.0003\,min period and a 95.2889\,$\pm$\,0.0005\,min period. The $\simeq$\,95.3\,min phase-fold shows two maxima and minima per cycle (Fig.\,\ref{f-WDJ1616_phasefold_g}), and we interpret this as the true period of variability. To obtain a more precise estimate of the period, we performed a $\chi^2$ fit using the sinusoidal function $\Delta\textrm{Mag} = A\sin(2 \pi t/P - \phi)$ around the $\simeq$\,95.3\,min value, where $t$ is the time of each $\Delta$Mag measurement, $P$ is the period, $A$ is the amplitude and $\phi$ is the phase zero-point. This resulted in a period solution of 95.288828\,$\pm$\,0.000023\,min which we provide as our final solution. The variations seen in the ZTF and LT $g$-band data behave in a similar manner, whereas the ZTF $r$-band data appear to vary at a slightly reduced amplitude; a colour dependence on the strength of variability that has been previously observed at DAHe white dwarfs \citep{redingetal20-1}.

\begin{figure}
\centerline{\includegraphics[width=1\columnwidth]{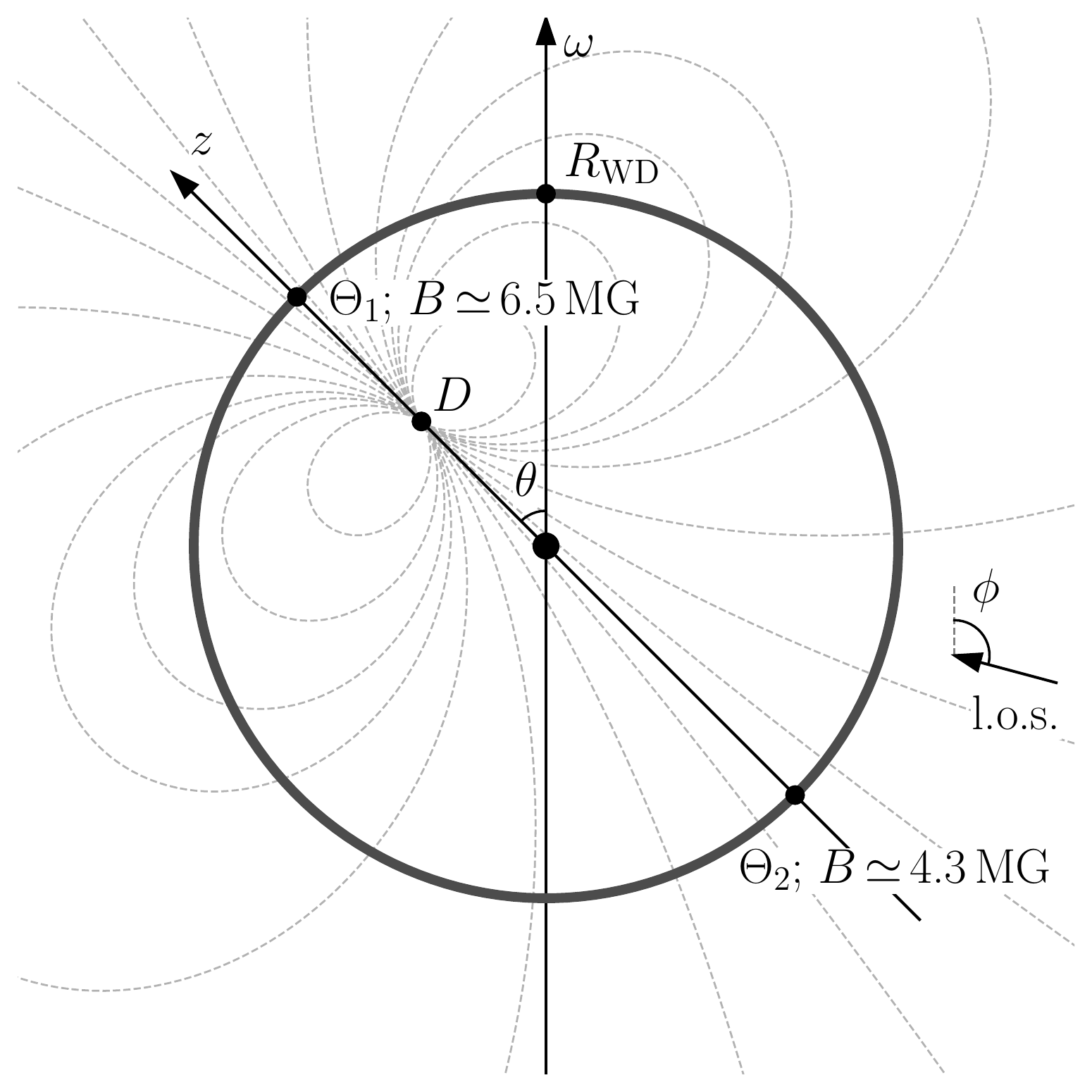}}
\caption{\label{f-toymodel} A toy model of the field geometry for an offset magnetic dipole at WD\,J1616+5410. The magnetic axis, $z$, is misaligned by the angle, $\theta$, with respect to the rotational axis, $\omega$. The centre of the dipolar field, $D$, is offset from the centre of the white dwarf with radius $R_\mathrm{WD}$. The two magnetic poles of the white dwarf are labelled as $\Theta_1$ and $\Theta_2$, where $\Theta_1$ has a higher field strength as it is located closer to the centre of the dipole. An observer sees the system along the line of sight (l.o.s.) which is at an angle, $\phi$, from the rotational axis. As the white dwarf rotates, the magnetic distribution and the visibility of the magnetic poles to the observer will vary.}
\end{figure}

With the multi-epoch DESI spectroscopy showing variations in the strength of both the observed magnetic field and Balmer emission, we convert the shutter open time and exposure duration from Table\,\ref{t-spectral_dates} to phase space which are shown in Fig.\,\ref{f-WDJ1616_phasefold_g}. It is immediately clear that spectra \#3 and \#12 are offset in by $\simeq$\,0.5 in phase, with spectrum \#3 (\#12) aligned with the deepest (shallowest) minima in the photometry at phase $\simeq$\,0.2 (0.7). These observations are consistent with those seen at all confirmed DAHe white dwarfs \citep{ferrarioetal97-1,brinkworthetal04-1, gansickeetal20-1, redingetal20-1, waltersetal21-1}. For WD\,J1616+5410, we suggest an offset dipole field configuration with the two magnetic poles that rotate in and out of view (see Fig.\,\ref{f-toymodel} for a toy model), both of which have co-located, temperature-inverted, optically-thin hotspots. These magnetic poles have differing magnetic field strengths, resulting in slightly different photometric minima when the poles are in view. When viewing the system equator-on, we expect the white dwarf to be at photometric maximum and its spectrum to present the lowest field strengths with Zeeman-split \textit{absorption}, as the poles are no longer in full view at the limbs of the observable hemisphere (Fig.\,\ref{f-toymodel}). This is consistent with the DESI spectra and LT and ZTF photometry, where spectra \#2, \#4, and \#6 each align with one of the two photometric maxima in Fig.\,\ref{f-WDJ1616_phasefold_g}.

Other epochs of spectroscopy have poor $S/N$, or show featureless (\#5) or potentially blended emission and absorption features (\#13). The relatively long exposure times ($\gtrsim$\,13\,min) of the DESI spectroscopy are likely to blame for the blended or featureless profiles, and the poor $S/N$ achieved for spectrum \#7 in a short exposure show the limits achievable with serendipitously obtained, survey-based observations. Follow-up observations at a higher $S/N$ and shorter exposure times are required to further analyse the spectroscopic variability seen at WD\,J1616+5410.

\section{Identification of 21 additional DAHe systems}\label{s4}

\subsection{DESI identified DAHe systems}

Motivated by the identification of WD\,J1616+5410, we expanded our search for DAHe white dwarfs in the DESI EDR and DR1 samples which include observations of over 47\,000 white dwarf candidates from \cite{fusilloetal19-1}. The Zeeman-split Balmer profiles can shift from their rest-wavelengths location across a significant fraction of the optical spectrum when magnetic field strengths reach several tens of MG (see fig.\,2 of \citealt{schmidtetal03-1}), making it difficult to select for these systems with simple criteria (e.g. equivalent width at the rest-wavelength of H$\alpha$). We therefore opted to visually inspect the entire sample searching for spectra with signs of emission features across the entire spectral range of DESI. In addition to WD\,J1616+5410, we identified 30 potential systems as well as two of the previously known DAHe systems; WD\,J1219+4715 and WD\,J1252--0234 \citep{gansickeetal20-1, redingetal20-1}. We attempted to identify magnetic field strengths for the 30 candidate DAHe white dwarfs in a manner similar to what was done for WD\,J1616+5410. We were able to confirm 16 of the candidates as new DAHe by matching their emission features with Zeeman-split Balmer line transitions (Table\,\ref{t-DAHe_systems}). For most of the systems we only had a single DESI spectrum, with six systems having two, three, or four spectra, and hence we could obtain no information on the field topology. For completeness we also list the 14 systems where a magnetic field could not be associated with the emission features as DESI DAHe candidates.

\subsection{SDSS identified DAHe systems}\label{s4SDSS}

Eight of the DAHe systems identified from their DESI spectra were previously observed by the SDSS and classified as DC white dwarfs \citep{mccook+sion99-1, harrisetal03-1, eisensteinetal06-1, kleinmanetal13-1, kepleretal15-1, kepleretal16-1, fusilloetal19-1, mccleeryetal20-1}, in addition to the previously published system WD\,J1219+4715 \citep{gansickeetal20-1}. In five of these systems (see Table\,\ref{t-DAHe_systems}) there is evidence of Zeeman-split Balmer line emission in the SDSS spectra, although it is not surprising these subtle features were missed as the class of DAHe white dwarfs has only recently been established. In fact, GD\,356, the prototypical system was originally classified as a DC white dwarf \citep{greenstein74-1,ferrarioetal97-1}. This motivated us to visually re-inspect the all the 2621 SDSS spectra of white dwarfs classified as DCs (see Section\,\ref{s2SDSS}), and discovered an additional four DAHe systems: WD\,J011027.51--102008.82, WD\,J101949.67+114148.73, WD\,J122619.77+183634.46\footnote{\cite{kawka+vennes06-1} noted that WD\,J122619.77+183634.46, also known as LP\,435--109, hosted H\,$\alpha$ emission and suggested the system may be a close binary, but classified the system as a DC.}, and WD\,J143657.48+210714.67 (hereafter WD\,J0110--1020, WD\,J1019+1141, WD\,J1226+1836, and WD\,J1436+2107 respectively). We were able to measure magnetic field strengths for all four new DAHe systems (Table\,\ref{t-DAHe_systems}). 

The DESI and SDSS spectra of all 20 new DAHe (in addition to WD\,J1616+5410) are shown in Fig.\,\ref{f-normplot_appendix}, where each spectrum is normalised at the $B$ value that best matches the emission features.

\subsection{Periodic signals from ZTF data}

We searched the ZTF data of each of the additional 20 DAHe white dwarfs and 14 candidates for periodic signals in a similar manner to the analysis of WD\,J1616+5410. We identified significant photometric signals for eight of the DAHe systems and one of the candidates, which are reported in Table\,\ref{t-DAHe_systems}. We associate most of these signals with the spin period of the white dwarf, although in some cases ambiguities between more than one photometric signal remain, and we recommend dedicated photometric follow-up to corroborate the periods detected in the ZTF data~--~ideally following a strategy similar to the one we used for WD\,J1616+5410, which proved very efficient. We were not able to identify significant periodic signals for 12 of the 21 new DAHe systems reported here, including WD\,J1226+1836 which is a relatively bright system with $G=16.2$ within the volume-limited 40\,pc sample of white dwarfs \citep{tremblayetal20-1, mccleeryetal20-1}. This strongly suggests that photometric variability based searches for DAHe white dwarfs will be biased and incomplete. We propose that an unbiased, high $S/N$ spectroscopic survey of DC-identified white dwarfs in the region of the \textit{Gaia} HRD identified by Equations\,\ref{eq:hrdcut} will provide a more complete sample of DAHe systems.

\begin{table*}
\centering
\caption{DAHe white dwarfs observed by DESI and SDSS. Emboldened values are taken from the literature \citep{gansickeetal20-1,redingetal20-1}, and for WD\,J1616+5410 we give the highest field strength observed from the multi-epoch observations. The last column indicates whether emission is visible in SDSS spectroscopy. Systems with no SDSS spectra are marked with the ``-'' symbol. Periods marked $^\dag$ may be half the true white dwarf spin-period in case that two poles contribute to the observed photometric modulation, and those marked $^*$ are tentative. Two significant signals were identified for WD\,J0110--1020 and we provide both here. We also provide 14 low confidence candidates identified from DESI spectroscopy where magnetic field strengths could not be determined. \label{t-DAHe_systems}}
\begin{tabular}{lrrrr}
\hline
DESI Systems                            & $G$   & $B$ (MG) & $P$ (h)      & Emission identifiable in SDSS spectrum? \\
\hline                                                                
WD\,J125230.93--023417.72         & 17.46 & \textbf{5.0}  & \textbf{0.09} & -   \\   
WD\,J161634.36+541011.51          & 18.24 & 6.5           & 1.59          & -   \\
WD\,J144626.01+282600.21          & 18.22 & 7.3           & 0.75$^*$      & no  \\
WD\,J175611.14+335230.75          & 17.36 & 13.6          &               & -   \\
WD\,J145207.19+325240.45          & 19.26 & 14.2          &               & yes \\
WD\,J155807.89+381649.61          & 19.79 & 18.4          &               & -   \\
WD\,J121929.50+471522.94          & 17.53 & \textbf{18.5} & \textbf{15.26}& yes \\ 
WD\,J082337.16+383816.50          & 19.12 & 18.7          &               & no  \\
WD\,J150057.83+484002.41          & 18.65 & 19.0          & 1.42          & no  \\
WD\,J075224.17+472422.44          & 18.53 & 21.0          &  1.21$^\dag$  & yes \\
WD\,J140916.34--000011.32         & 18.64 & 23.5          &               & yes \\
WD\,J041926.91--011333.28         & 17.81 & 34.0          & 1.65          & -   \\
WD\,J171101.52+654549.87          & 18.74 & 37.7          &               & yes \\
WD\,J001319.16+240111.02          & 19.34 & 42.3          &               & -   \\
WD\,J002844.20+055931.27          & 18.48 & 45.2          &               & -   \\
WD\,J075429.35+661106.64          & 16.72 & 56.1          & 1.37          & -   \\
WD\,J073227.97+662310.18          & 17.58 & 99.1          & 34.3          & -   \\
WD\,J063357.87+561413.12          & 17.93 & 116           &               & -   \\
WD\,J132035.85+324925.05          & 18.60 & 147           &               & yes \\
\hline                                          
SDSS systems                      &       &               &               &     \\
\hline
WD\,J011027.51--102008.82         & 17.52 & 9.3           & 6.63 / 9.16   & yes \\
WD\,J101949.67+114148.73          & 18.88 & 10.5          & $0.44^*$      & yes \\
WD\,J122619.77+183634.46          & 16.23 & 11.5          &               & yes \\
WD\,J143657.48+210714.67          & 17.22 & 60.0          &               & yes \\
\hline                                         
DESI Candidates                   &       &               &               &     \\
\hline                          
WD\,J004736.08--154326.57         & 17.99 &               &               & -   \\
WD\,J012915.76--152430.04         & 18.51 &               &               & -   \\
WD\,J031921.71--035432.96         & 18.38 &               &               & -   \\
WD\,J041321.81--061330.41         & 19.65 &               &               & -   \\
WD\,J043042.29+003025.69          & 19.78 &               &               & -   \\
WD\,J071203.03+440710.94          & 19.33 &               &               & -   \\
WD\,J073547.95+682836.51          & 18.18 &               &               & -   \\
WD\,J102242.40+330550.42          & 18.70 &               &               & -   \\
WD\,J112444.53+564204.17          & 18.90 &               &               & no  \\
WD\,J124333.86+031737.13          & 18.58 &               &               & no  \\
WD\,J125940.74+623449.82          & 17.76 &               &               & no  \\
WD\,J143740.61+030900.68          & 17.57 &               &               & -   \\
WD\,J165132.13+374410.33          & 18.92 &               &               & no  \\
WD\,J173835.47+144120.91          & 18.69 &               & 8.96$^*$      & -   \\                
\hline
\end{tabular}
\end{table*}

\section{Discussion}\label{s5}

\subsection{Occurrence rates of DAHe white dwarfs}

\begin{figure*}
\centerline{\includegraphics[width=2\columnwidth]{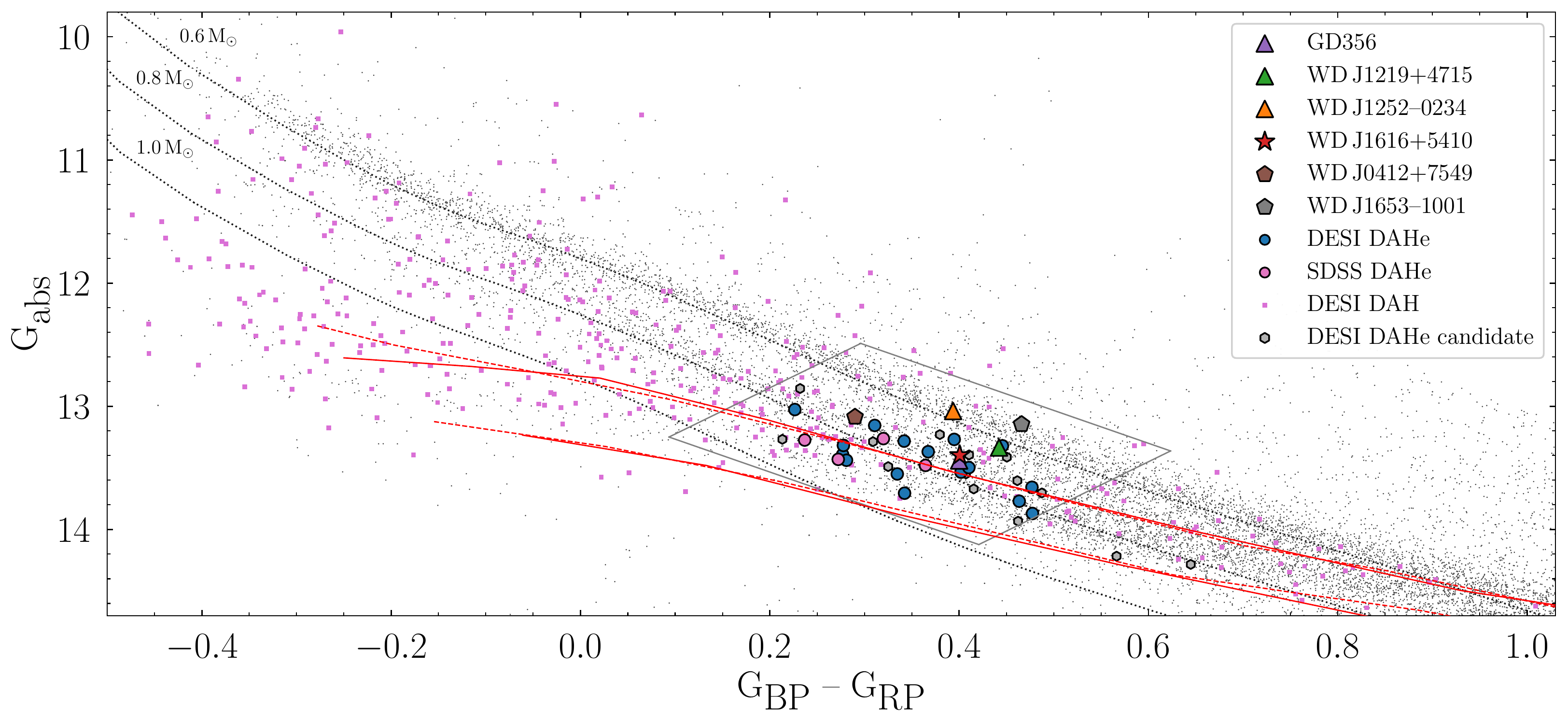}}
\caption{\label{f-GaiaHR} \textit{Gaia} HR diagram showing the three literature DAHe systems (triangles), two DAe systems (pentagons), WD\,J1616+5410 (star), the additional 20 DAHe systems identified in this paper (circles) and the 14 DAHe candidates (hexagons) (Table\,\ref{t-DAHe_systems}). White dwarf candidates within 100\,pc (gray dots, \protect{\citealt{fusilloetal19-1}}), and DAH white dwarfs identified in the DESI EDR and DR1 samples (pink squares) are also plotted. The gray box shows the region bound by Equations\,\ref{eq:hrdcut}. The black dotted lines show the cooling tracks of 0.6\,\Msun, 0.8\,\Msun, and 1.0\,\Msun\ DA white dwarfs \citep{bedardetal20-1}. The red solid and dashed lines show the crystallization sequences (which are a function of white dwarf mass) for thick- and thin-hydrogen atmosphere models, respectively, where 20\,per\,cent (upper) and 80\,per\,cent (lower) of the white dwarf has crystallized.}
\end{figure*}

We show the white dwarf cooling sequence in the \textit{Gaia} HRD in Fig.\,\ref{f-GaiaHR}, highlighting the location of the three previously published DAHe systems (WD\,J1252--0234, GD356, and WD\,J1219+4715), the two DAe white dwarfs (which show Balmer-line emission but no Zeeman splitting) WD\,J0412+7549 \citep{tremblayetal20-1, waltersetal21-1} and WD\,J1653--1001 \citep{obrienetal23-1}, WD\,J1616+5410 and our additional 20 DAHe white dwarfs, as well as the 14 candidate DAHe systems. The clustering of the DAHe systems on the white dwarf cooling track is striking, and suggests that their presence in this confined parameter space is related to an evolutionary process which occurs at late ($\gtrsim$\,1\,Gyr) cooling times. Previous studies have already commented on the tight clustering of the DAHe systems on the \textit{Gaia} HR diagram \citep{gansickeetal20-1, waltersetal21-1}, with \cite{schreiberetal21-2} suggesting that this is partially due to the onset of crystallization. 

To estimate the occurrence rate of these DAHe systems within the location they cluster in, we constructed an area defined by the set of equations, 

\begin{align}
\label{eq:hrdcut}
G_{\textrm{abs}} + \frac{15}{4}\left(G_{\textrm{BP}} - G_{\textrm{RP}}\right) - 13.6 & > 0 \\
G_{\textrm{abs}} - \frac{8}{3}\left(G_{\textrm{BP}} - G_{\textrm{RP}}\right) - 11.7  & > 0 \notag \\
G_{\textrm{abs}} + \frac{15}{4}\left(G_{\textrm{BP}} - G_{\textrm{RP}}\right) - 15.7  & <  0 \notag \\
G_{\textrm{abs}} - \frac{8}{3}\left(G_{\textrm{BP}} - G_{\textrm{RP}}\right) - 13.0  & <  0 \notag
\end{align}

\noindent which is depicted in Fig.\,\ref{f-GaiaHR} (we did not include the 14 DAHe candidates to constrain this region). We visually inspected the DESI EDR and DR1 spectra of 5155 white dwarf candidates contained within this area, 
and identified 5056 single white dwarfs, 19 white dwarfs in binaries\footnote{This includes WD\,J154905.35+193132.60, a newly-identified AM\,CVn.}, 21 extra-galactic sources (e.g. quasars), 18 stellar spectra, and 41 objects for which the spectroscopy was too poor to determine their nature. Of the 5056 isolated white dwarfs, we recover the 17 new DAHe white dwarfs identified here in addition to two of the previously known systrems, WD\,J1252--0234 and WD\,J1219+4715 (Table\,\ref{t-DAHe_systems}). This results an occurrence rate of $O_{\mathrm{DAHe/WD}}=0.38\pm0.09$\,per\,cent for white dwarfs in the area defined by Equations\,\ref{eq:hrdcut}
exhibiting observable Zeeman-split Balmer lines in emission, with uncertainties determined by sampling from a binomial distribution.

From our visual inspection of the $\simeq$\,47\,000 DESI EDR and DR1 white dwarf candidates we also identified 368 magnetic DA white dwarfs without Balmer line emission (DAHs)\footnote{Full information on these targets will be released in a future paper.}, of which 72 are present in the boxed region (Fig.\,\ref{f-GaiaHR}). Using the same method as above, we determine an occurrence rate of DAHs without emission in the boxed region as $O_{\mathrm{DAH/WD}}=1.4\pm0.2$\,per\,cent. Taking the total number of magnetic DA white dwarfs in the box (72 DAH plus 19 DAHe systems), $O_{\mathrm{DAHe/(DAH+DAHe)}}=21\pm4$\,per\,cent are DAHe systems. 

$O_{\mathrm{DAHe/WD}}$ and $O_{\mathrm{DAH/WD}}$ are strict lower limits for DAHe and DAH systems given $S/N$ constraints on identifying weak features, as well as possible phase-dependent emission that are not seen in the DESI exposures (e.g. spectrum \#4 in Fig.\,\ref{f-3_4} would be identified as a DAH rather than a DAHe). $O_{\mathrm{DAH/WD}}$ is also additionally effected by the ability to detect fields using the presence of Zeeman-splitting in Balmer absorption features below $\simeq$\,1\,MG. 

To factor in the bias of detecting faint emission features at higher $S/N$ values, we recalculated the above occurrence rates for a series of sub-samples which had a lower limit set on $S/N$ in the range observed for the DAHe systems ($17.8 \leq S/N \leq 140$). The number of white dwarf spectra in these sub-samples decreased from 1884 to ten. The occurrence rates are reasonably consistent over a range of $30 \leq S/N \leq 90$, with median values and standard deviations of $O_{\mathrm{DAHe/WD}}=1.5\pm0.3$\,per\,cent, $O_{\mathrm{DAH/WD}}=3.4\pm0.5$\,per\,cent, and $O_{\mathrm{DAHe/(DAH+DAHe)}}=30\pm5$\,per\,cent. We take these as the occurrence rates of DAH and DAHe from the DESI sample, although they are still likely lower limits (Table\,\ref{t-occ_rates}). 

To compare against a more complete sample, we calculated the same occurrence rate of DAH and DAHe systems to isolated white dwarfs in the 40\,pc volume-limited sample of white dwarfs \citep{tremblayetal20-1,mccleeryetal20-1, obrienetal23-1} contained in the boxed region in Fig.\,\ref{f-GaiaHR}. The 40\,pc sample benefits from many (but not all) white dwarfs having dedicated observations with higher $S/N$ compared with DESI spectroscopy. The region of interest contains 168 isolated white dwarfs of which thirteen are DAH systems, resulting in an occurrence rate of $O_{\mathrm{DAH/WD40pc}}=8\pm2$\,per\,cent. $O_{\mathrm{DAH/WD40pc}}$ and $O_{\mathrm{DAH/WD}}$ are in rough agreement ($\simeq$\,2.3$\sigma$ difference), although the 40\,pc sample has a larger occurrence rate which we attribute to the increased $S/N$ and spectral resolution of that sample, allowing weaker fields (i.e. narrower Zeeman splitting) to be probed. WD\,J1226+1836, identified here as a new DAHe system from SDSS spectra, is the second DAHe found in the 40\,pc sample after GD\,356 (WD\,J0412+7549 and WD\,J1653--1001 are DAe white dwarfs) and we find $O_{\mathrm{DAHe/WD40pc}}=1.3\pm_{0.7}^{1.1}$\,per\,cent which is in excellent agreement with $O_{\mathrm{DAHe/WD}}$. Finally, we obtain an occurrence rate of DAHe systems compared to all magnetic white dwarfs (DAHe\,+\,DAH) in the 40\,pc sample in the boxed region as $O_{\mathrm{DAHe/(DAH+DAHe)40pc}}=14\pm_{7}^{10}$\,per\,cent. The 1:1 ratio of DAe:DAHe systems in the 40\,pc sample is tentatively suggestive of an unidentified population of DAe systems in the spectroscopic white dwarf samples from DESI and SDSS, but identifying these is beyond the scope of this study. A summary of the occurrence rates calculated for the DESI and 40\,pc samples are given in Table\,\ref{t-occ_rates}.

\begin{table}
\centering
\caption{Occurrence rates, $O_{X/Y}$, of systems $X$ compared to sample of $Y$ systems for the DESI and 40\,pc samples in the area on the \textit{Gaia} HRD described by equations\,\ref{eq:hrdcut}
(see text and Fig.\,\ref{f-GaiaHR}). DESI occurrence rates are determined using spectra with signal-to-noise ratios in the range $30 \leq S/N \leq 90$. \label{t-occ_rates}} 
\begin{tabular}{lrr}
\hline
Occurrence rate & DESI (per\,cent) & 40\,pc (per\,cent) \\
\hline                                  
$O_{\mathrm{DAHe/WD}}$               & 1.5\,$\pm$\,0.3 & 1.3\,$\pm$\,$_{0.7}^{1.1}$  \\
$O_{\mathrm{DAH/WD}}$                & 3.4\,$\pm$\,0.5 & 8\,$\pm$\,2                 \\
$O_{\mathrm{DAHe/(DAH+DAHe)}}$       & 30\,$\pm$\,5    & 14\,$\pm$\,$_{7}^{10}$        \\
\hline
\end{tabular}
\end{table}

The occurrence rates calculated above suggest that $\simeq$\,10\,--\,30\,per\,cent of DAH systems in the boxed region show emission features, implying DAHe may be far more common than previous searches have indicated. 
These DAHe white dwarfs likely either form from DAH white dwarfs, where a mechanism begins to produce spectral emission features, or they form from DAs, where both the magnetic field and emission features emerge together. If DAH white dwarfs are the progenitors for DAHe systems, the origin of the emission is not necessarily a phase that all DAH white dwarfs go through, or it is a phase that is not long lived. If instead a subset of DAs transition into DAHe white dwarfs, then this channel could potentially provide up to a $\simeq$\,45\,per\,cent increase in magnetic hydrogen-atmosphere white dwarfs over this region of the cooling track.  \cite{schreiberetal21-2} recently suggested that a rotational and crystallization driven dynamo in a DA white dwarf enhanced by the accretion of planetary material could explain the presence of DAHe white dwarfs.

\subsection{Discussion on the origin of DAHe systems}

\subsubsection{Can the late onset of DAHe white dwarfs in the \textit{Gaia} HRD be explained by the production of a magnetic field from a crystallization-driven convective dynamo?}

The stellar spin-up resulting from the accretion of planetary material \citep{stephanetal20-1} along with the onset of crystallization setting up a convective dynamo offers a potentially plausible scenario for the appearance of magnetic Balmer emission features at a specific time in the cooling sequence of isolated white dwarfs \citep{isernetal17-1, ginzburgetal22-1}. However, recent studies have brought into question whether the convective velocities in crystallizing white dwarfs are large enough to explain the observed magnetic field strengths \citep{fuentesetal23-1}. Here we focus on whether the observational evidence is in agreement with a crystallization-driven dynamo generating the magnetic fields seen in DAHe systems.

Spin-up of a white dwarf and the generation of a magnetic field has been previously linked to planetary bodies through the engulfment of sub-stellar bodies in a common envelope during the red/asymptotic giant branch phases along with a common envelope driven dynamo \citep{siess+livio99-1, farihietal11-2}. The presence of these bodies are expected, and four such giant sub-stellar candidates that have survived the giant branch phases have already been identified around white dwarfs \citep{gaensickeetal19-1, vanderburgetal20-1, blackmanetal21-1,gaia22-1}. While planetary engulfment during a common envelope could explain the more rapid spin-periods and presence of a magnetic field for DAHe white dwarfs, the generation of a magnetic field during the common envelope cannot explain the the appearance of DAHe systems at late times in the \textit{Gaia} HRD.

To examine the scenario that the onset of DAHe systems is in-part due to crystallization, we used the evolutionary models of \cite{bedardetal20-1}\footnote{Cooling models available at \url{http://www.astro.umontreal.ca/~bergeron/CoolingModels}} to determine the crystallization tracks for thick hydrogen atmospheres (${M_{\textrm{H}}}/{M_{\textrm{WD}}}$\,=\,10$^{-4}$, where $M_{\textrm{H}}$ is the mass of hydrogen in the white dwarf atmosphere and $M_{\textrm{WD}}$ is the white dwarf mass) and thin hydrogen atmospheres (${M_{\textrm{H}}}/{M_{\textrm{WD}}}$\,=\,10$^{-10}$). These crystallization tracks are plotted in Fig.\,\ref{f-GaiaHR} with upper and lower bounds denoting 20\,per\,cent and 80\,per\,cent of the white dwarf mass being crystallized respectively, as presented in \cite{tremblayetal19-1}.

\begin{figure}
\centerline{\includegraphics[width=1\columnwidth]{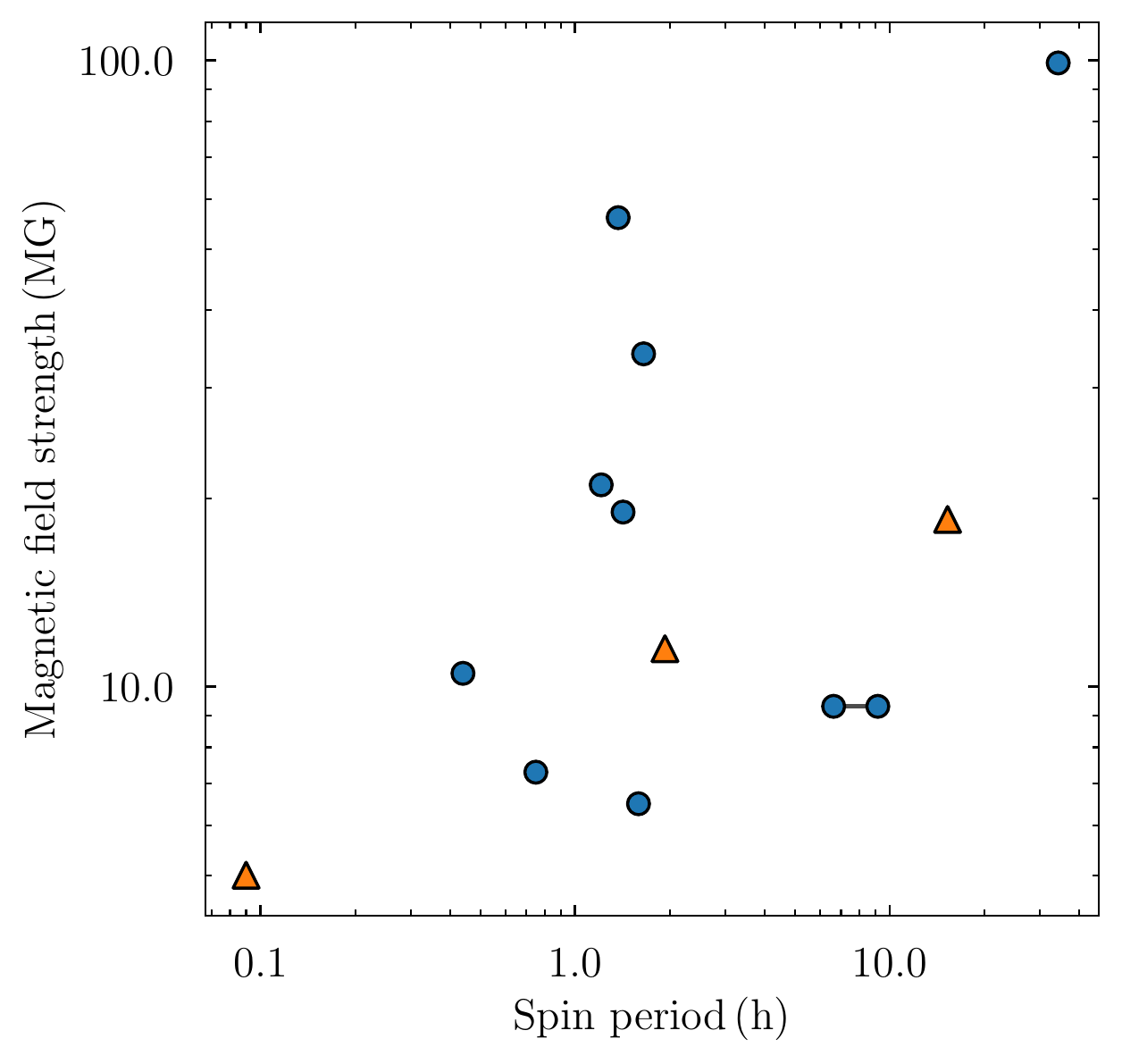}}
\caption{\label{f-BvsP} Magnetic field strength against spin period for DAHe white dwarfs from literature (orange triangles, \citealt{gansickeetal20-1,redingetal20-1,waltersetal21-1}) and this paper (blue circles). Two significant periodic signals were identified for WD\,J0110--1020, which are connected by a gray line. A slight positive trend is seen, which is not in agreement with the relationship $B$\,$\propto$\,$P^{-1/2}$ determined by \protect\cite{ginzburgetal22-1} for a magnetic field generated by a crystallization-driven convective dynamo.}
\end{figure}

Out of the 24\footnote{After submission of this manuscript, a preprint was posted announcing the discovery of two additional DAHe \citep{redingetal23-1}, bringing the number of currently known DAHe to 26.} known DAHe systems, only ten fall within the region at which crystallization sets in. Furthermore, for a crystallization-driven dynamo one would expect an anti-correlation between the magnetic field strength and the spin-period of the white dwarf (see \citealt{ginzburgetal22-1} and their fig.\,4). However, the three DAHe systems identified in the literature, along with the nine DAHe systems discovered here with ZTF-determined periodic signals show a positive correlation between $B$ and $P$ (Fig.\,\ref{f-BvsP}). This discrepancy was noted by \cite{ginzburgetal22-1} for the three previously known DAHe systems, and our new additions suggest an even steeper correlation between $P$ and $B$. A caveat is that the strongest periodic signals identified from ZTF photometry may be off by a factor of two from the true rotation period (as it is the case for WD\,J1616+5410), however, that systematic uncertainty does not change the apparent correlation seen in Fig.\,\ref{f-BvsP}. Photometric follow-up of the new DAHe systems presented here will eventually allow the dependence between magnetic field strength and white dwarf spin-period to be more rigorously tested.

If a crystallization and rotation driven dynamo creates the magnetic fields in these systems, then they should evolve and cool as a standard DA white dwarf until they reach the crystallization track and become magnetic. The distribution of DAHe systems appear to more closely match the distribution of DAH white dwarfs in the boxed region of the \textit{Gaia} HRD (which are thought to have masses that clusters around $M_{\textrm{WD}}$\,=\,0.8\,\Msun\ identified from the magnitude limited sample of SDSS white dwarfs, \citealt{ferrarioetal20-1}), rather than the canonical DA cooling track for $M_{\textrm{WD}}$\,=\,0.6\,\Msun. This would suggest that the progenitors of DAHe systems are more likely to be similar to DAHs rather than DAs. However, it is also likely that these DAHe systems have radiative atmospheres, as convection should be inhibited by the presence of a strong magnetic field \citep{gentilefusilloetal18-1, gansickeetal20-1}. The resulting differences in the structures of their atmospheres can alter their spectral energy distribution and shift their location on the \textit{Gaia} HRD compared to a standard, non-magnetic, DA white dwarf. Mismatches between $T_{\textrm{eff}}$ and spectral energy distributions determined from ultraviolet and optical spectra have also been observed for magnetic white dwarfs \citep{schmidtetal86-2, gaensickeetal01-1}. In conclusion, the currently available data and theoretical models cannot unambiguously constrain whether crystallization can ubiquitously explain the emergence of DAHe white dwarfs. Further work on these systems, in particular measuring their spin-periods and understanding their location on the \textit{Gaia} HRD are needed to investigate the crystallization-driven dynamo as a universal origin mechanism for the DAHe systems.

\subsubsection{What is the generation mechanism for emission in DAHe white dwarfs?}

Whether or not a crystallization-driven convective dynamo can explain the onset of a magnetic field in these DAHe systems, an additional mechanism is needed to produce the observed Zeeman-split emission features. Observations show that these emission features are produced by optically thin hot-spots co-located with regions on the white dwarf surface that are darker in the continuum. The unipolar inductor model, whereby a close-in conductive body heats up the white dwarf atmosphere, has been invoked to explain the temperature inverted hot-spot creating the emission feature seen in spectroscopy \citep{goldreich+lyndenbell69-1, lietal98-1, wickramasingheetal10-1}. The model has also been used to explain the apparent cutoff in the \textit{Gaia} HRD of DAHe systems by a natural lifetime put in place on the conducting planetary body through Lorentz drift \citep{veras+wolszczan19-1, gansickeetal20-1}. However, \cite{waltersetal21-1} argue that the rapid stellar rotation periods inhibit current carriers from reaching the white dwarf and therefore heating the surface. Additionally, these white dwarfs do not exhibit metal absorption features indicative of the accretion of planetary material \citep{koesteretal97-1, jura03-1}. However, an absence of photospheric metal absorption features has been observed in the magnetic white dwarfs in polar-type cataclysmic variables \citep{gaensickeetal06-2}, which are definitively actively accreting. The lack of photospheric metals in polars is explained by the fact that the strong magnetic fields inhibit horizontal spreading of material before it sinks below the photosphere. In contrast, non-magnetic cataclysmic variables do all show clearly detectable metal features in their spectra \citep{sionetal95-3, gaensicke+beuermann96-1, palaetal17-1}. Furthermore, as these DAHe white dwarfs may have radiative atmospheres, horizontal spreading would also be suppressed when compared to white dwarfs with convective atmospheres \citep{cunninghametal21-1}, so the lack of photospheric metal absorption features is not necessarily indicative of a lack of accretion. 

Another scenario that has been suggested to explain the occurrence of DAHe white dwarfs is the presence of a chromosphere and chromospheric activity in these white dwarfs \citep{musielaketal05-1,waltersetal21-1}. Finally, \cite{bagnulo+landstreet22-1} observe an increase in the occurrence of magnetic white dwarfs in the volume-limited 40\,pc sample over the first 2\,-\,3\,Gyr, where the authors suggest these fields are left over from earlier stages in evolution and may emerge at later times. While this range roughly overlaps the onset of DAHe white dwarfs, and would potentially explain their delayed generation, further work is needed to explain the increase in the occurrence of magnetic fields with white dwarfs in time, and a possible link to DAHe white dwarfs.

\section{Conclusion}\label{s6}

Even though GD\,356, the first DAHe system, was discovered over 30 years ago \citep{greenstein+mccarthy85-1}, the physical origin of these systems remains uncertain. We provide detailed follow-up of a new DAHe with multi-epoch DESI spectroscopy, WD\,J1616+5410, where we identify photometric variability on a 95.3\,min period, and a magnetic field which varies between $3.5 < B < 6.5$\,MG. As it has been observed among the previously known DAHe white dwarfs, the brightness of WD\,J1616+5410 varies in anti-phase with the strength of the emission features and the observed average magnetic field strength across the observable hemisphere. This suggests that a photospheric dark spot, and an optically-thin temperature-inverted hot-spot are co-located close to or at the magnetic poles.

We have identified a sample of 21 DAHe systems from the DESI survey and the SDSS, providing a significant increase in the number of these enigmatic sources known and confirming their clustering at late cooling times on the \textit{Gaia} HRD. We calculate occurrence rates for these systems using the magnitude-limited DESI sample and the volume-limited 40\,pc sample of white dwarfs, which suggest that $\simeq$\,1\,per\,cent of white dwarfs are classed as DAHe in the region of the \textit{Gaia} HRD where they cluster, and $\simeq$\,10\,-\,30\,per\,cent of DAH white dwarfs within the same region exhibit Balmer line emission. Given ten DAHe systems are identifiable from archival SDSS spectroscopy, nine of which were previously classified as DC white dwarfs, an unbiased, high $S/N$ survey of DC white dwarfs in the boxed region of Fig.\,\ref{f-GaiaHR} may reveal more DAHe systems that are currently below the detection threshold. As only nine of our 21 DAHe systems have detectable periodic signals from ZTF data, searches based on photometric variability are likely to provide a biased sample of systems. Whether or not a crystallization and rotation driven dynamo can create DAHe white dwarfs is not distinguishable using the currently available data, and further theoretical work is required to confidently exclude this evolutionary channel. Spectroscopic and photometric follow-up of these newly identified DAHe white dwarfs will aid the understanding and characterisation of these systems.

\section*{Acknowledgements}

We thank the referee John Landstreet, whose insightful comments improved the quality of this manuscript. We thank Elena Cukanovaite for insightful discussions and Mairi O'Brien for sharing data on the 40\,pc volume-limited white dwarf sample. The authors acknowledge financial support from Imperial College London through an Imperial College Research Fellowship grant awarded to CJM. This project has received funding from the European Research Council (ERC) under the European Union’s Horizon 2020 research and innovation programme (Grant agreement No. 101020057). BTG was supported by the UK STFC grant ST/T000406/1. This research was supported in part by the National Science Foundation under Grant No. PHY-1748958.

Based on observations obtained with the Samuel Oschin Telescope 48-inch and the 60-inch Telescope at the Palomar
Observatory as part of the Zwicky Transient Facility project. ZTF is supported by the National Science Foundation under Grants
No. AST-1440341 and AST-2034437 and a collaboration including current partners Caltech, IPAC, the Weizmann Institute for
Science, the Oskar Klein Center at Stockholm University, the University of Maryland, Deutsches Elektronen-Synchrotron and
Humboldt University, the TANGO Consortium of Taiwan, the University of Wisconsin at Milwaukee, Trinity College Dublin,
Lawrence Livermore National Laboratories, IN2P3, University of Warwick, Ruhr University Bochum, Northwestern University and
former partners the University of Washington, Los Alamos National Laboratories, and Lawrence Berkeley National Laboratories.
Operations are conducted by COO, IPAC, and UW.

The Liverpool Telescope is operated on the island of La Palma by Liverpool John Moores University in the Spanish Observatorio del Roque de los Muchachos of the Instituto de Astrofisica de Canarias with financial support from the UK Science and Technology Facilities Council.

This research is supported by the Director, Office of Science, Office of High Energy Physics of the U.S. Department of Energy under Contract No. DE–AC02–05CH11231, and by the National Energy Research Scientific Computing Center, a DOE Office of Science User Facility under the same contract; additional support for DESI is provided by the U.S. National Science Foundation, Division of Astronomical Sciences under Contract No. AST-0950945 to the NSF’s National Optical-Infrared Astronomy Research Laboratory; the Science and Technologies Facilities Council of the United Kingdom; the Gordon and Betty Moore Foundation; the Heising-Simons Foundation; the French Alternative Energies and Atomic Energy Commission (CEA); the National Council of Science and Technology of Mexico (CONACYT); the Ministry of Science and Innovation of Spain (MICINN), and by the DESI Member Institutions: \url{https://www.desi.lbl.gov/collaborating-institutions}.

The authors are honored to be permitted to conduct scientific research on Iolkam Du’ag (Kitt Peak), a mountain with particular significance to the Tohono O’odham Nation.

\section*{Data availability}
Spectroscopy from the SDSS and photometry from ZTF are available from their respective public archives. Data presented here along with the Python scripts used to produce the figures in this manuscript are available at \url{https://zenodo.org/record/7581523#.Y9aAX1LP3Ah}.



\bibliographystyle{mnras}
\bibliography{aamnem99,aabib}




\appendix

\section{Multi-epoch spectroscopy of WD\,J1616+5410}

\begin{figure*}
\centerline{\includegraphics[width=2\columnwidth]{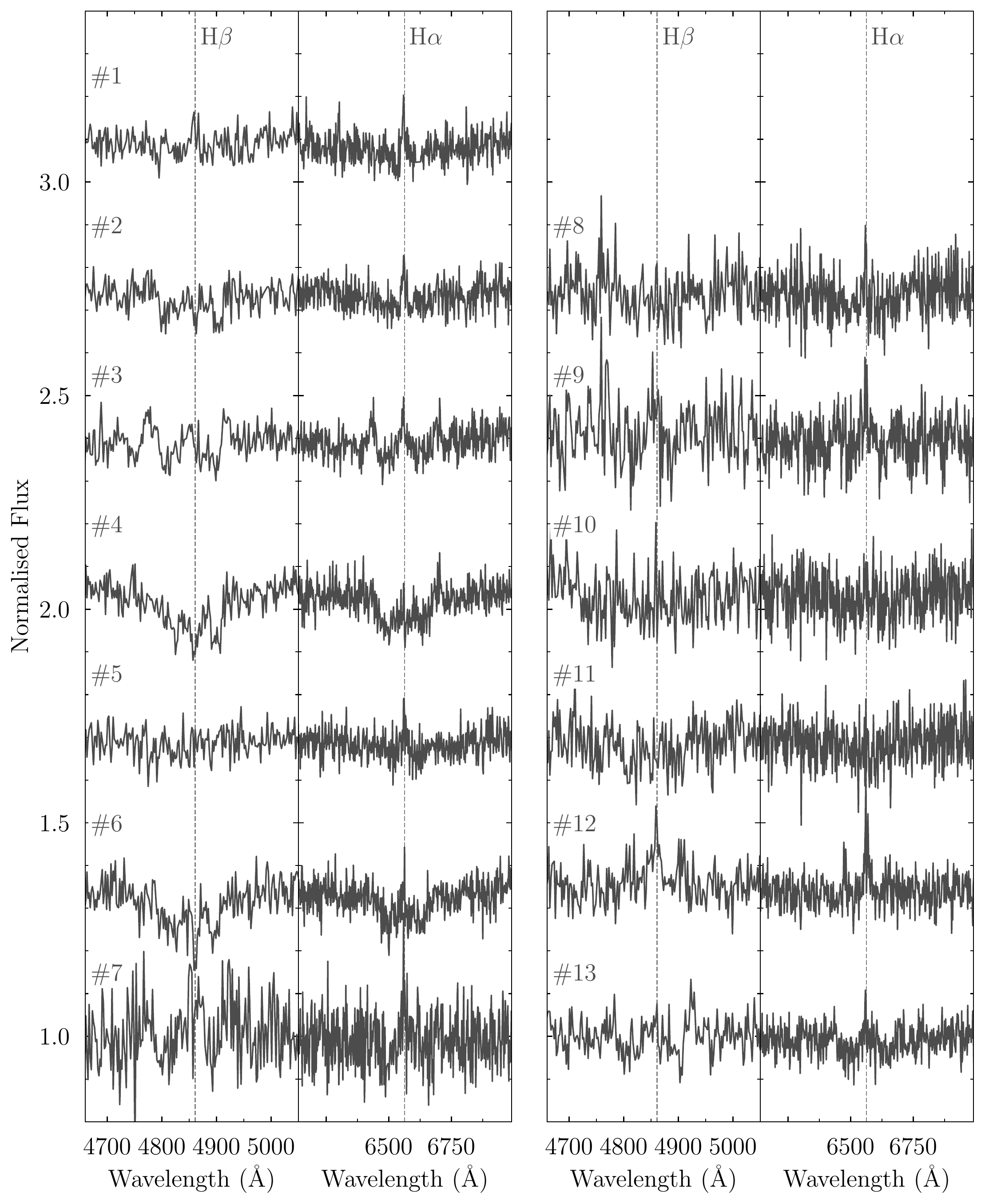}}
\caption{\label{f-all_spec} Continuum normalised DESI spectra of WD\,J1616+5410 showing the H$\beta$ and H$\alpha$ regions, which are offset in steps of 0.5 from 1.0 for clarity. The spectral numbers correspond to those in Table\,\ref{t-spectral_dates}, where \#3 and \#4 show the clearest variation of the Zeeman-split Balmer features.}
\end{figure*}

The 13 spectra obtained for WD\,J1616+5410 are presented here (Fig.\,\ref{f-all_spec}).

\section{Spectra of DAHe white dwarfs observed by DESI and SDSS}

\begin{figure*}
\centerline{\includegraphics[width=1.8\columnwidth]{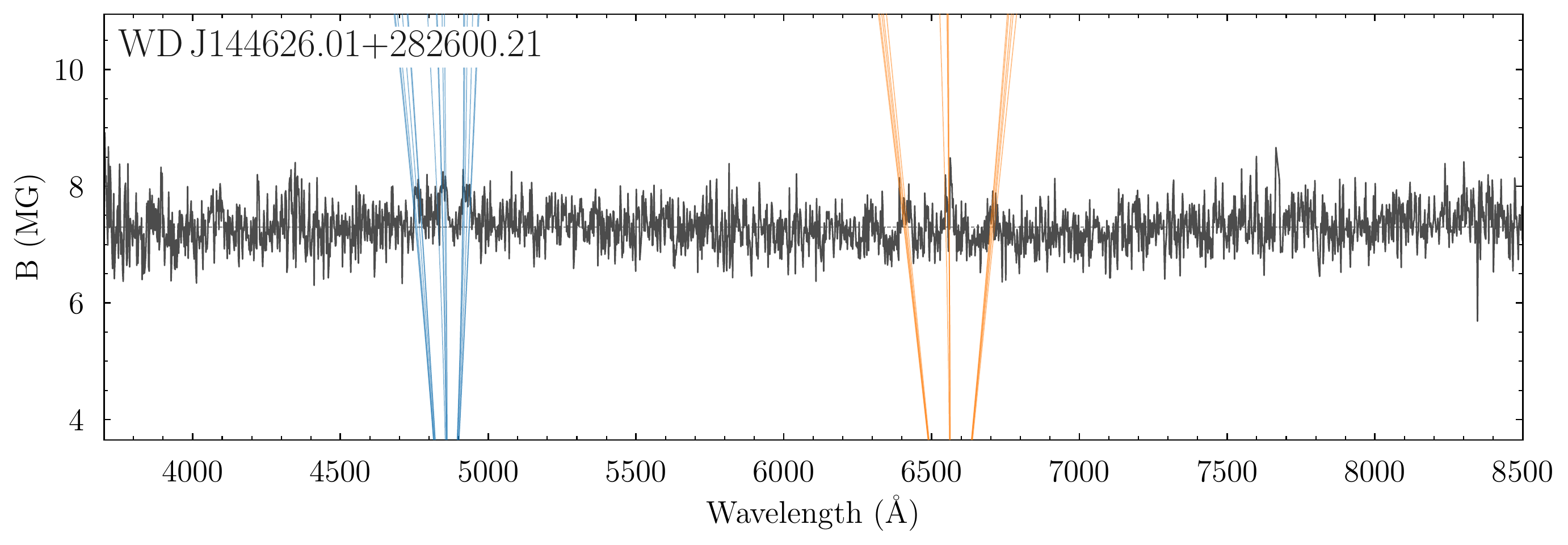}}
\centerline{\includegraphics[width=1.8\columnwidth]{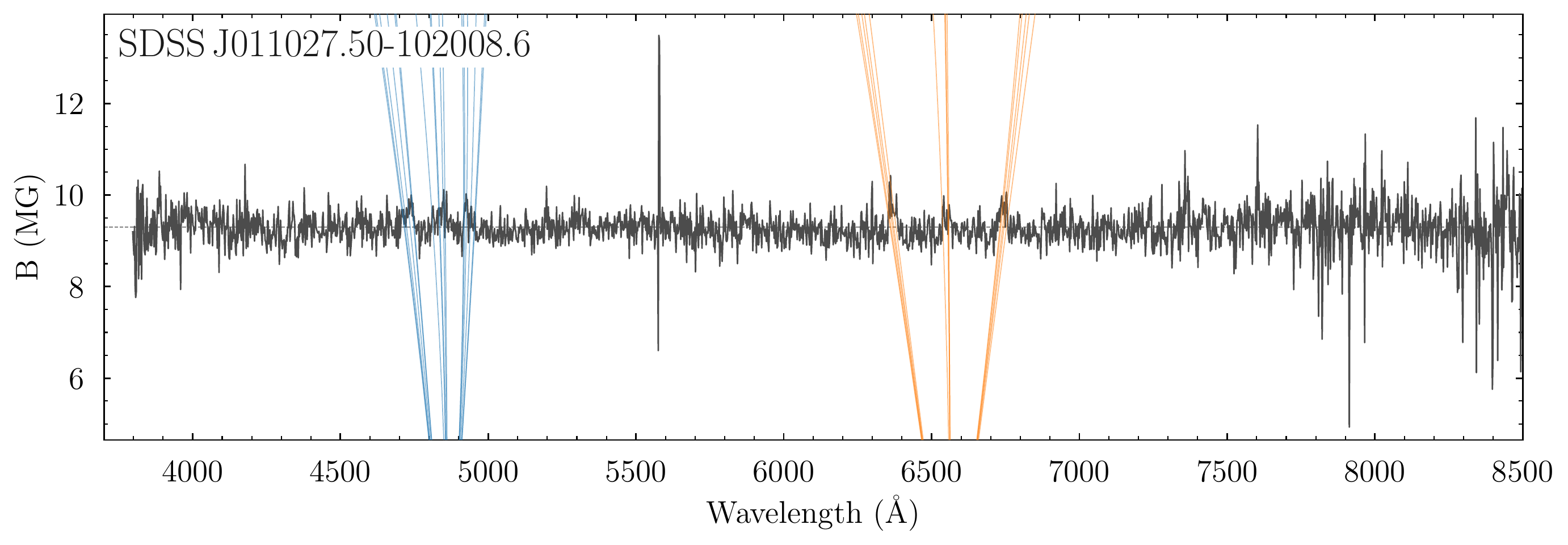}}
\centerline{\includegraphics[width=1.8\columnwidth]{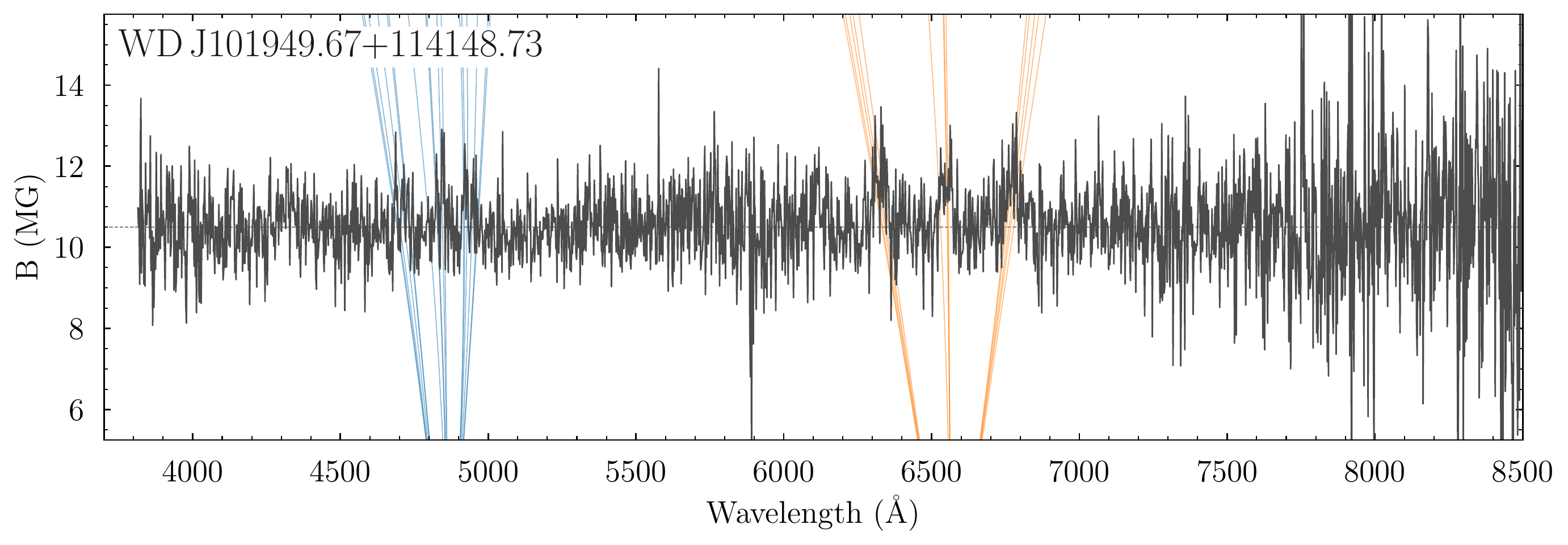}}
\centerline{\includegraphics[width=1.8\columnwidth]{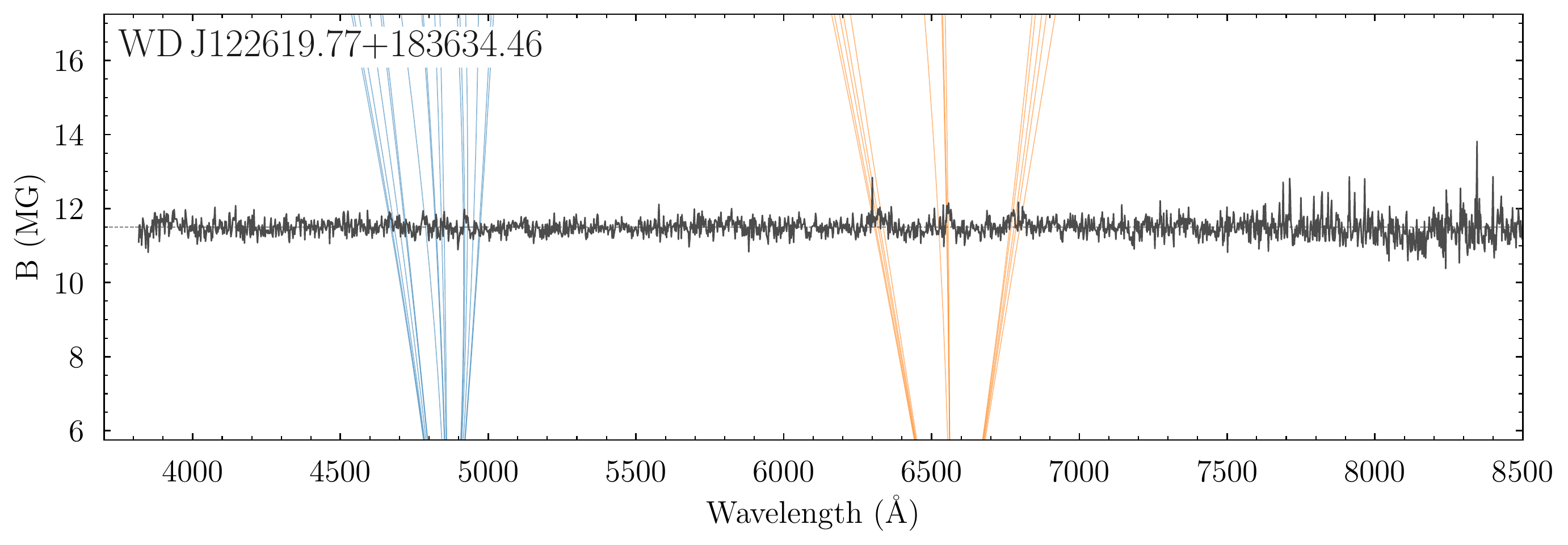}}
\caption{\label{f-normplot_appendix} Continuum normalised spectra of the newly-identified DAHe systems listed in Table\,\ref{t-DAHe_systems} excluding WD\,J1616+5410, where the normalised flux has been multiplied by the estimated magnetic field strength for each system. Transition wavelengths as a function of $B$ are plotted for the Zeeman-split components of H\,$\beta$ and H\,$\alpha$ in blue and orange respectively, and systems are plotted in order of increasing $B$.}
\end{figure*}

Provided here are the spectra of DAHe systems observed by DESI and SDSS (Table\,\ref{t-DAHe_systems}), with estimated field strengths (Fig.\,\ref{f-normplot_appendix}).

\renewcommand{\thefigure}{A\arabic{figure}}
\setcounter{figure}{0}

\begin{figure*}
\centerline{\includegraphics[width=1.8\columnwidth]{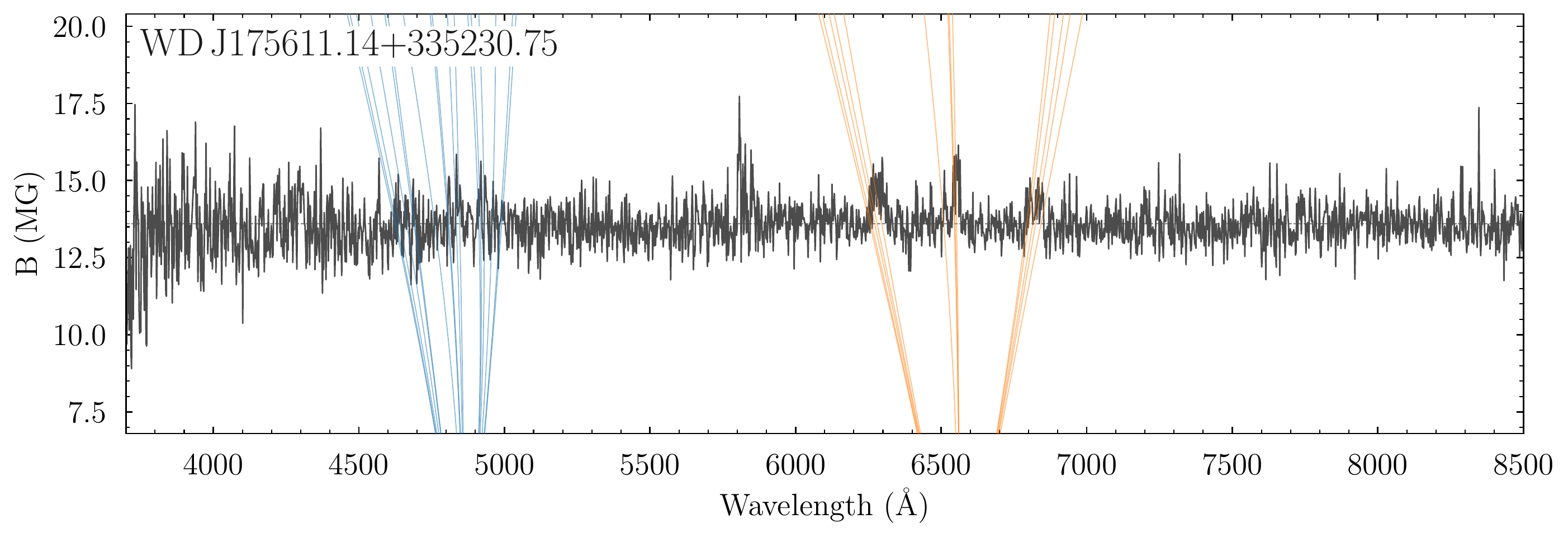}}
\centerline{\includegraphics[width=1.8\columnwidth]{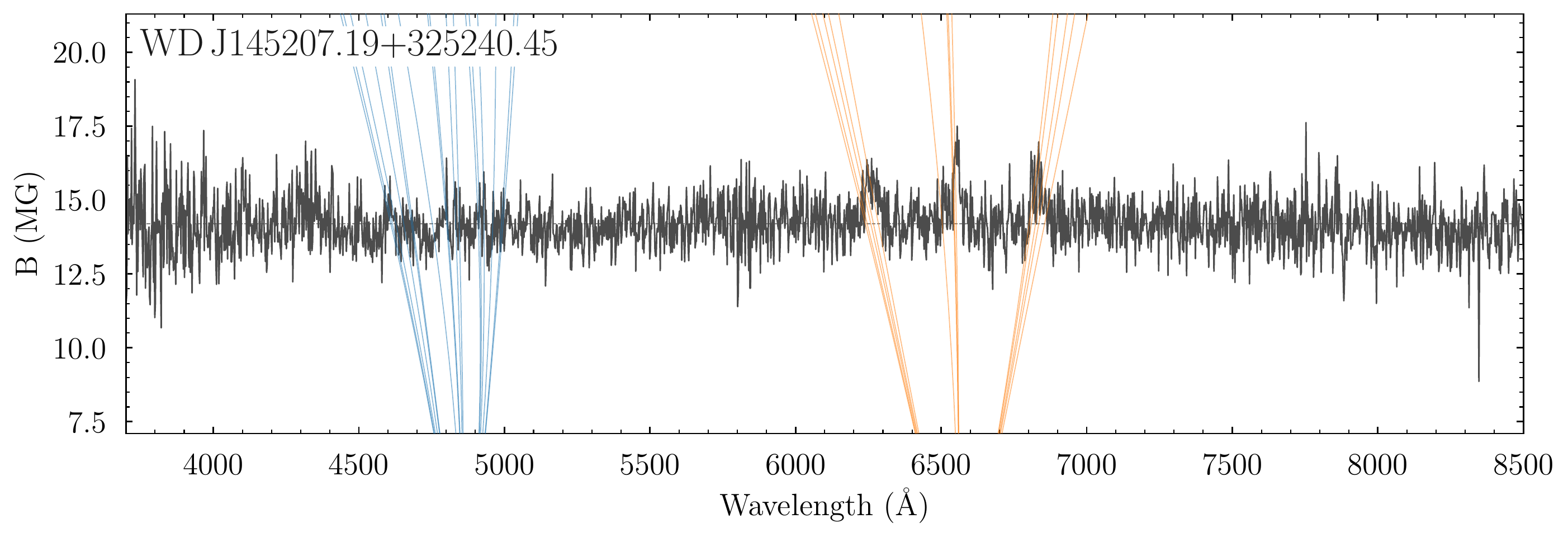}}
\centerline{\includegraphics[width=1.8\columnwidth]{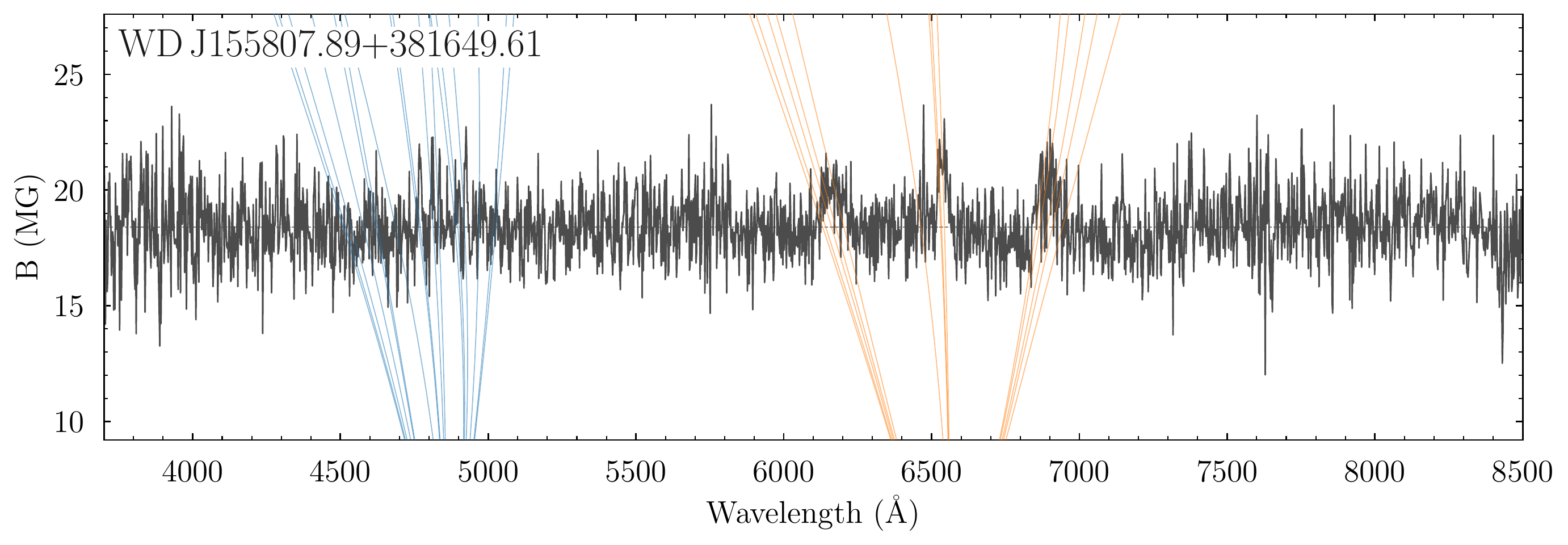}}
\centerline{\includegraphics[width=1.8\columnwidth]{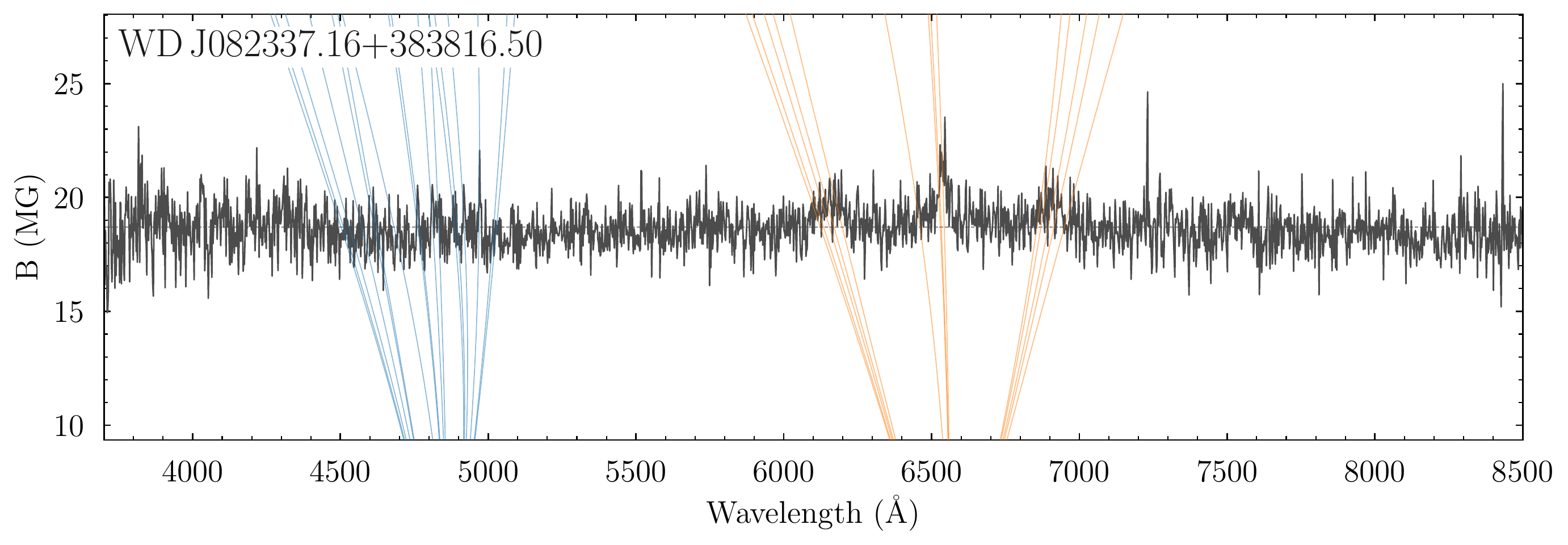}}
\caption{\label{f-normplot_appendix2} Continued.}
\end{figure*} 

\renewcommand{\thefigure}{A\arabic{figure}}
\setcounter{figure}{0}

\begin{figure*}
\centerline{\includegraphics[width=1.8\columnwidth]{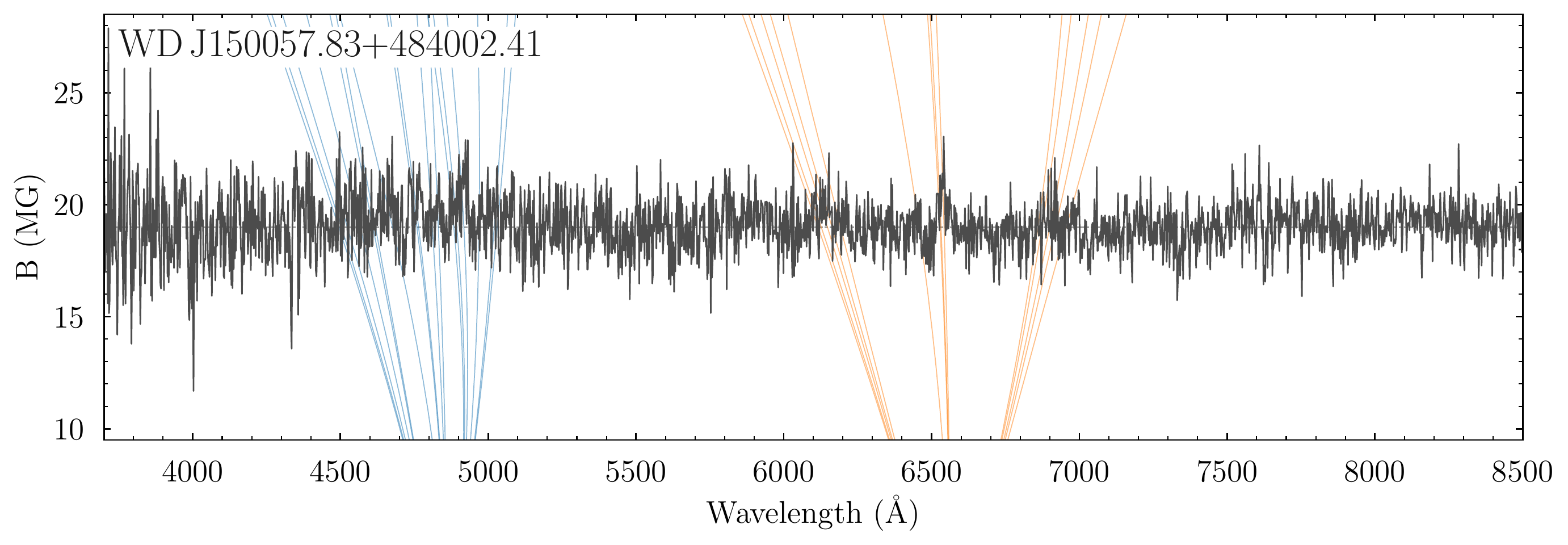}}
\centerline{\includegraphics[width=1.8\columnwidth]{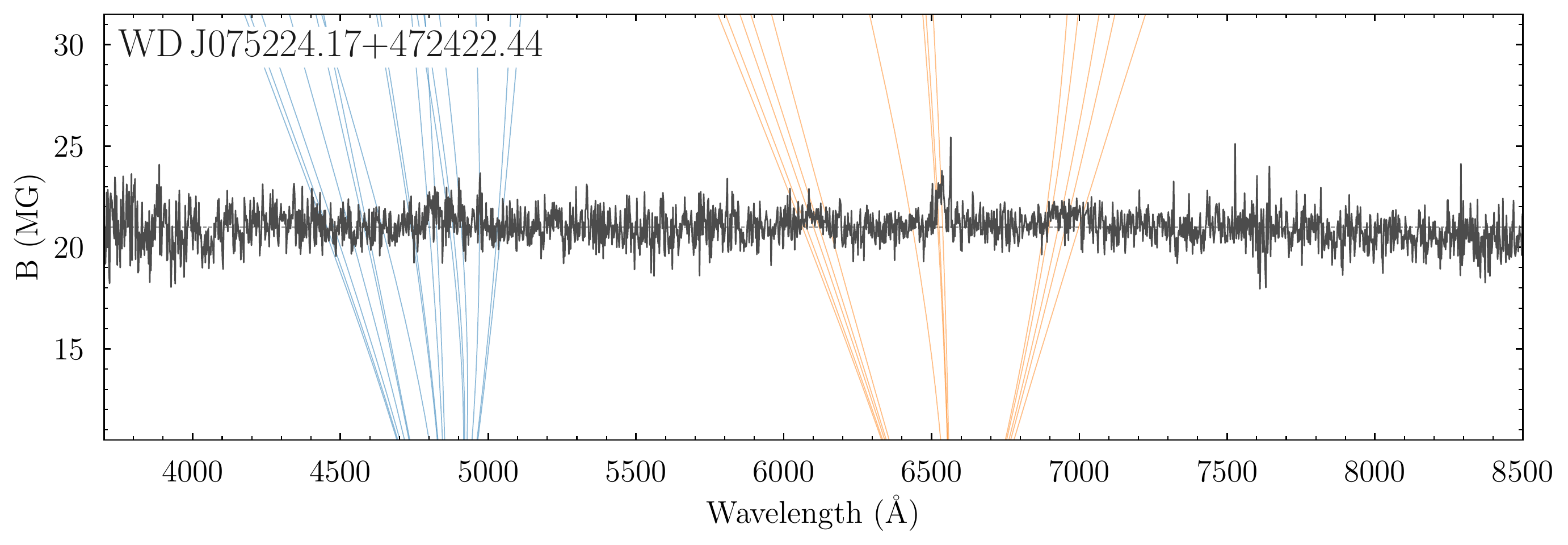}}
\centerline{\includegraphics[width=1.8\columnwidth]{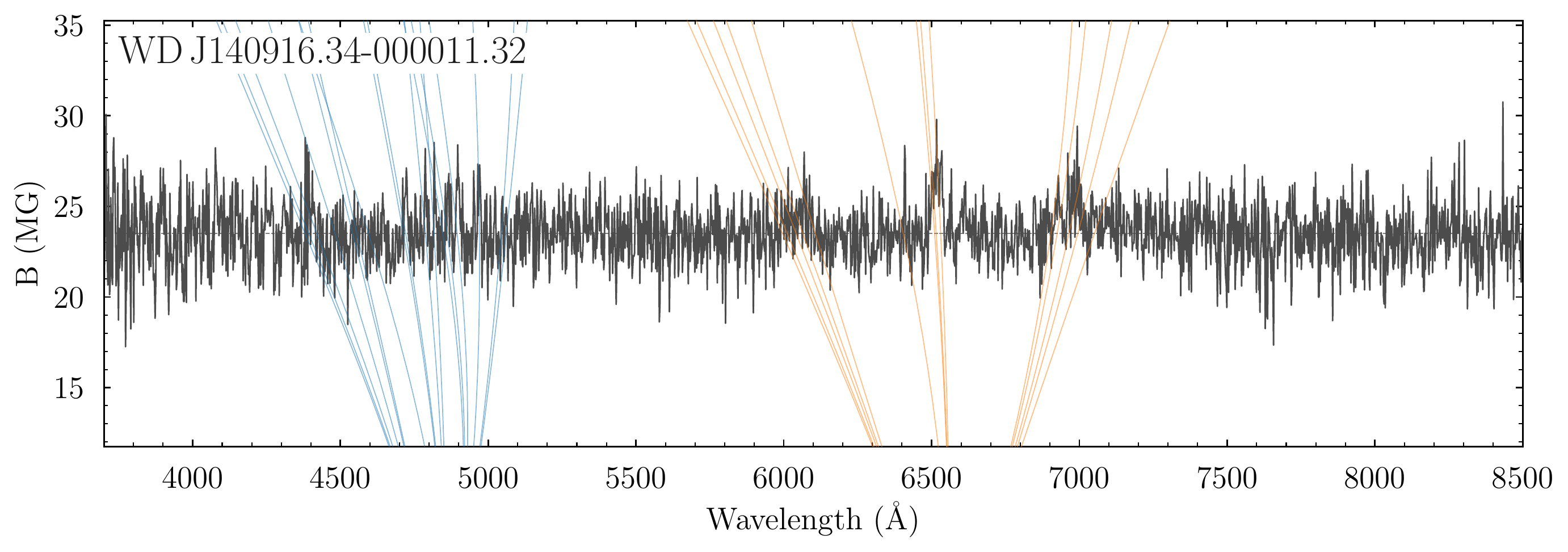}}
\centerline{\includegraphics[width=1.8\columnwidth]{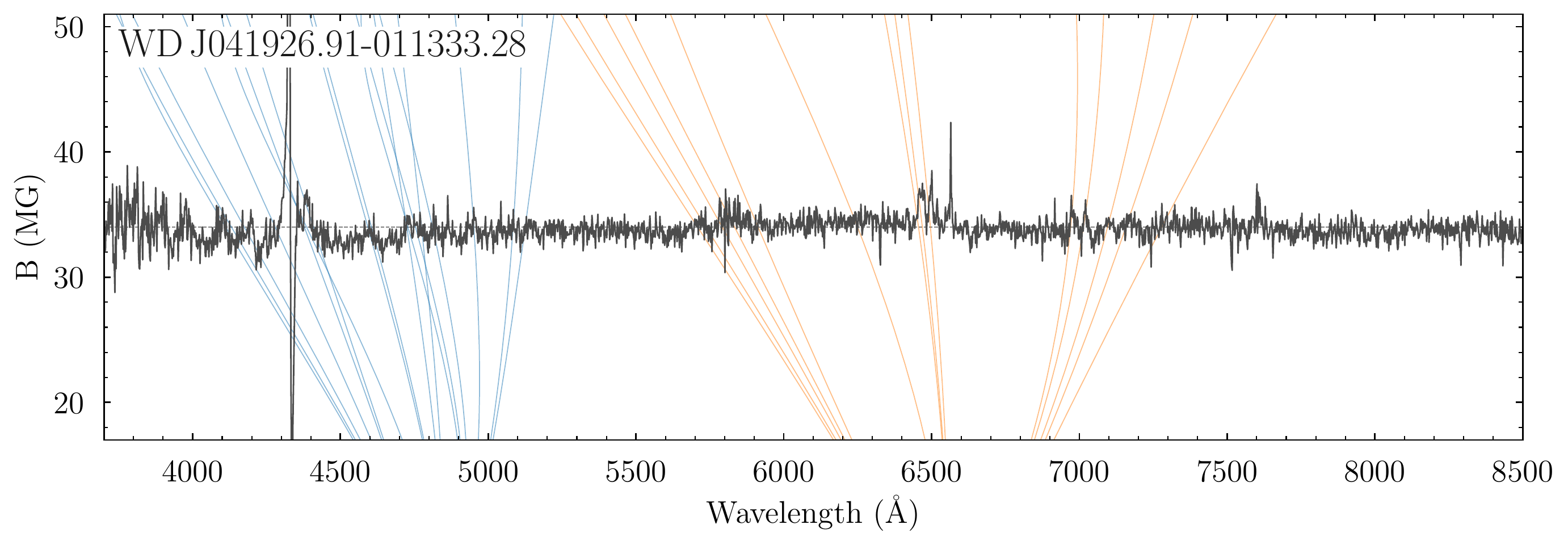}}

\caption{\label{f-normplot_appendix3} Continued.}
\end{figure*} 

\renewcommand{\thefigure}{A\arabic{figure}}
\setcounter{figure}{0}

\begin{figure*}
\centerline{\includegraphics[width=1.8\columnwidth]{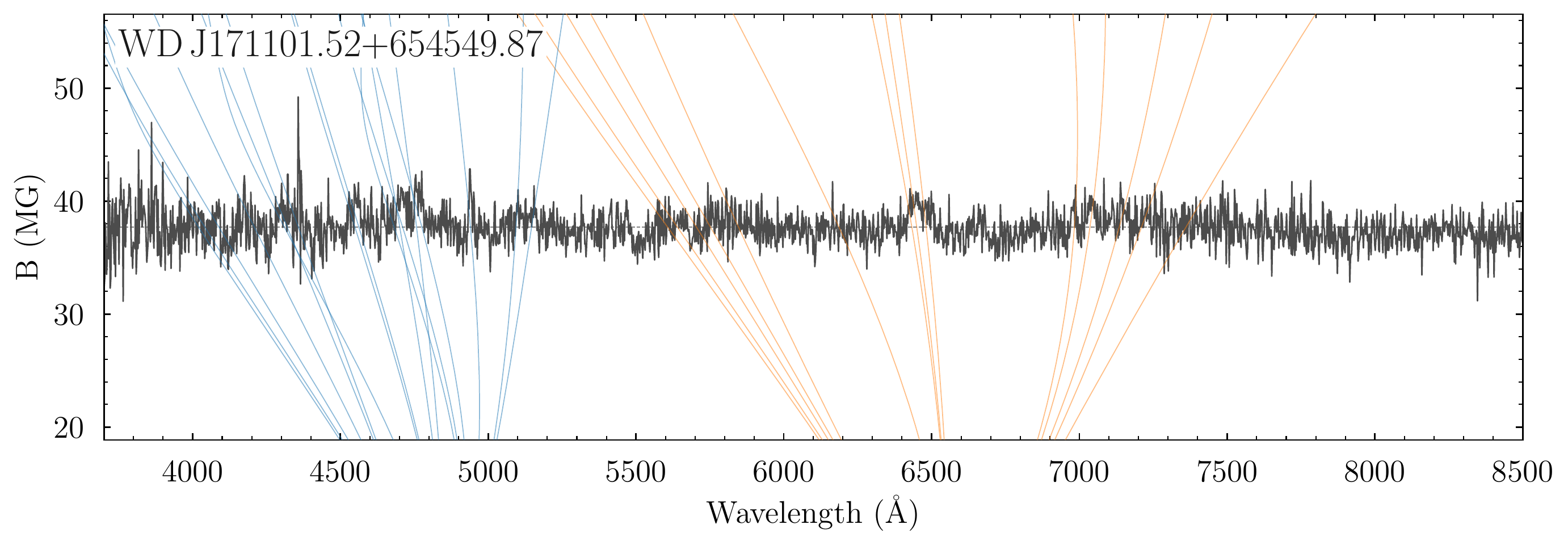}}
\centerline{\includegraphics[width=1.8\columnwidth]{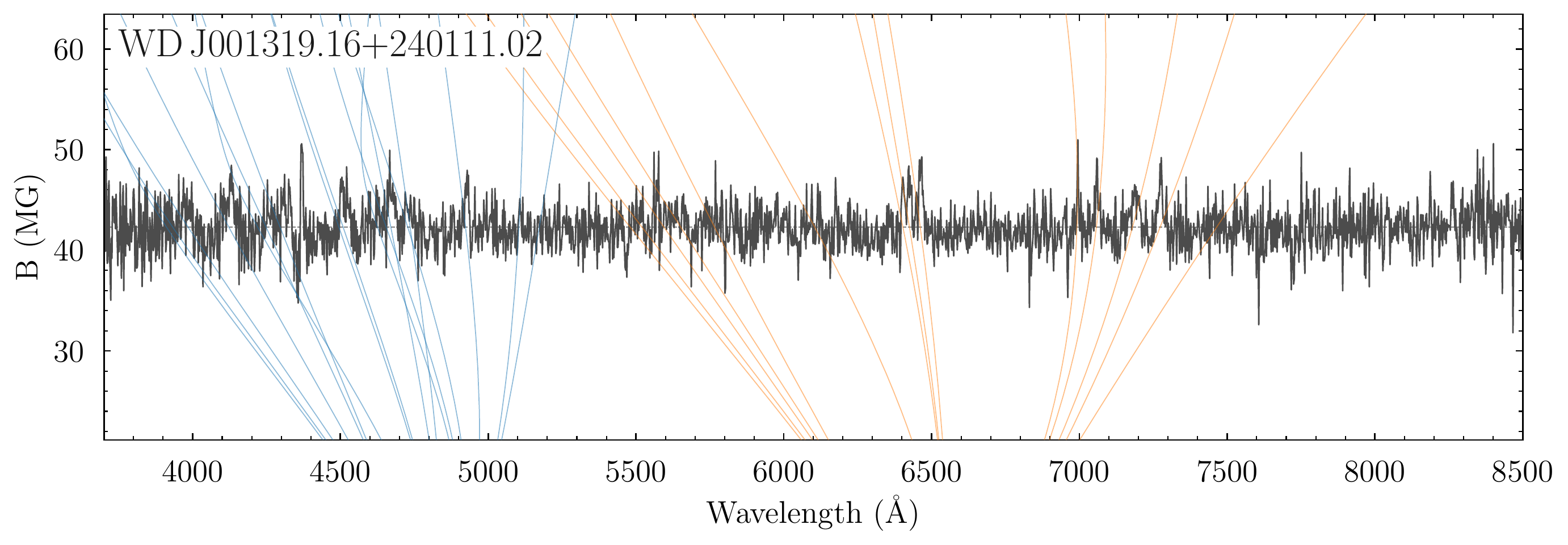}}
\centerline{\includegraphics[width=1.8\columnwidth]{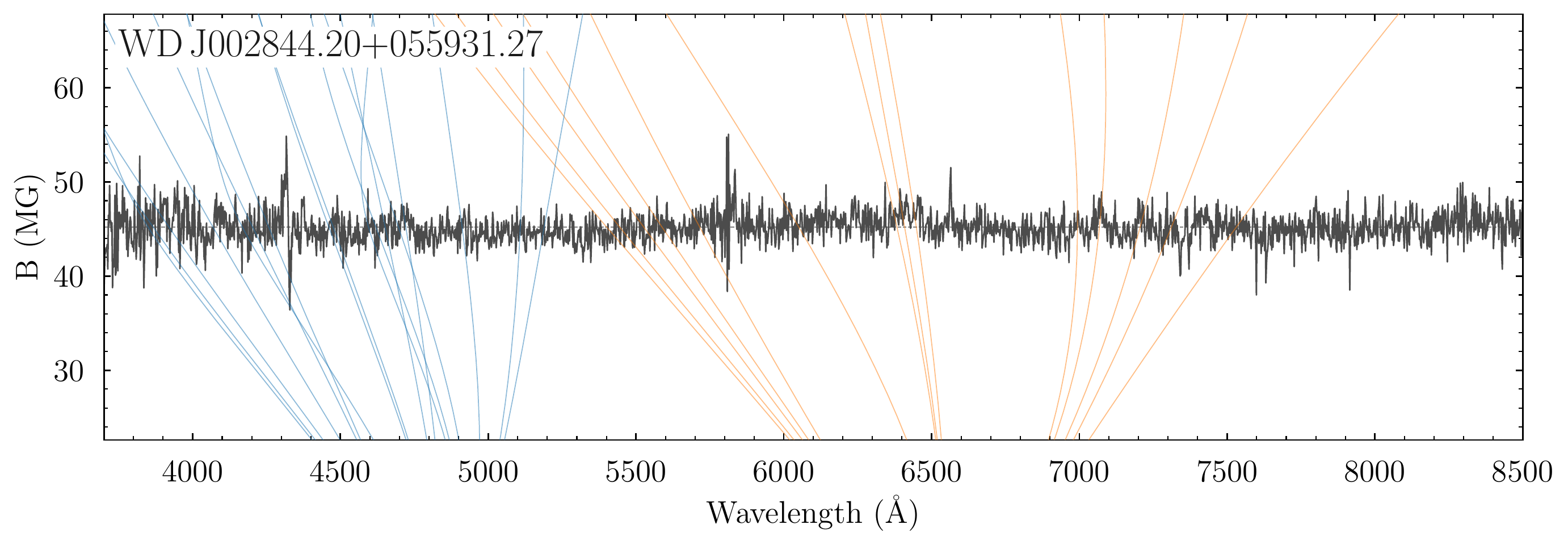}}
\centerline{\includegraphics[width=1.8\columnwidth]{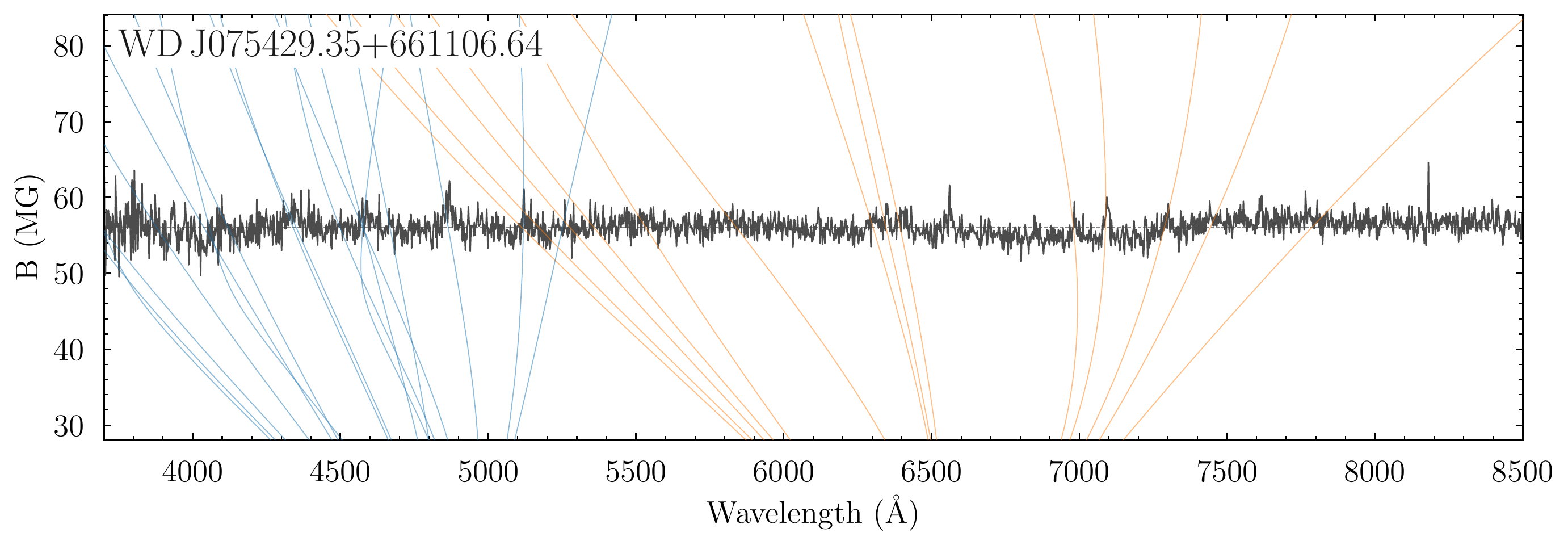}}

\caption{\label{f-normplot_appendix4} Continued.}
\end{figure*} 

\renewcommand{\thefigure}{A\arabic{figure}}
\setcounter{figure}{0}

\begin{figure*}
\centerline{\includegraphics[width=1.8\columnwidth]{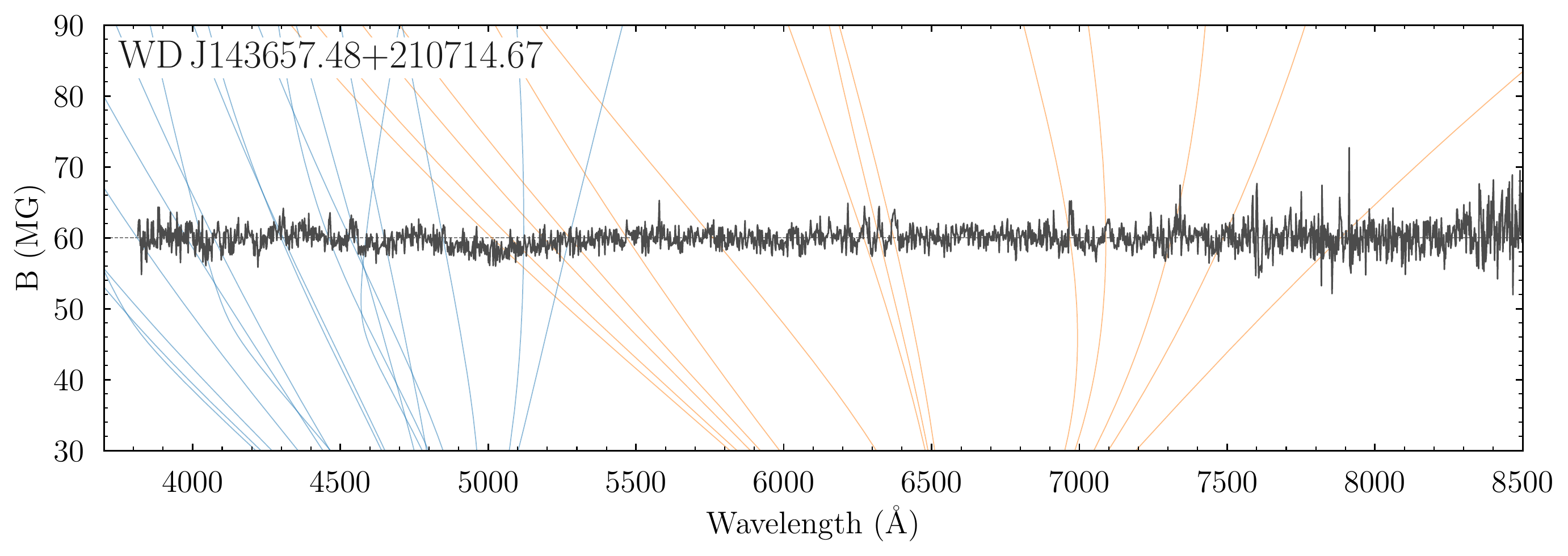}}
\centerline{\includegraphics[width=1.8\columnwidth]{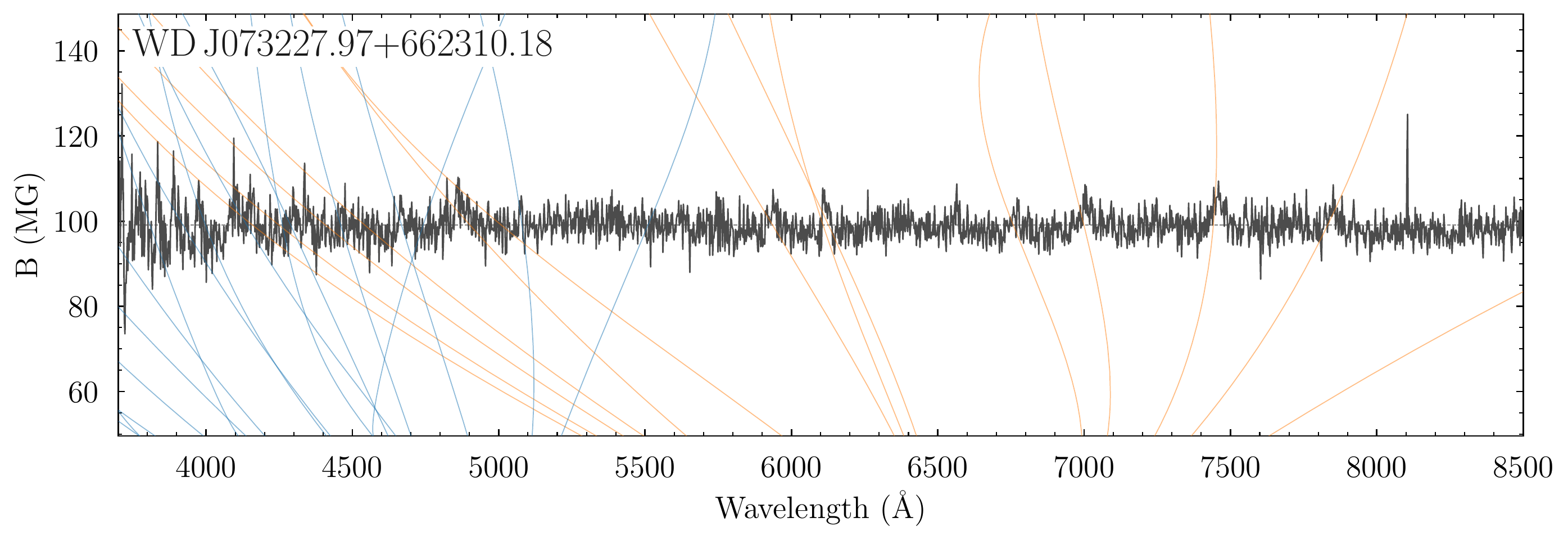}}
\centerline{\includegraphics[width=1.8\columnwidth]{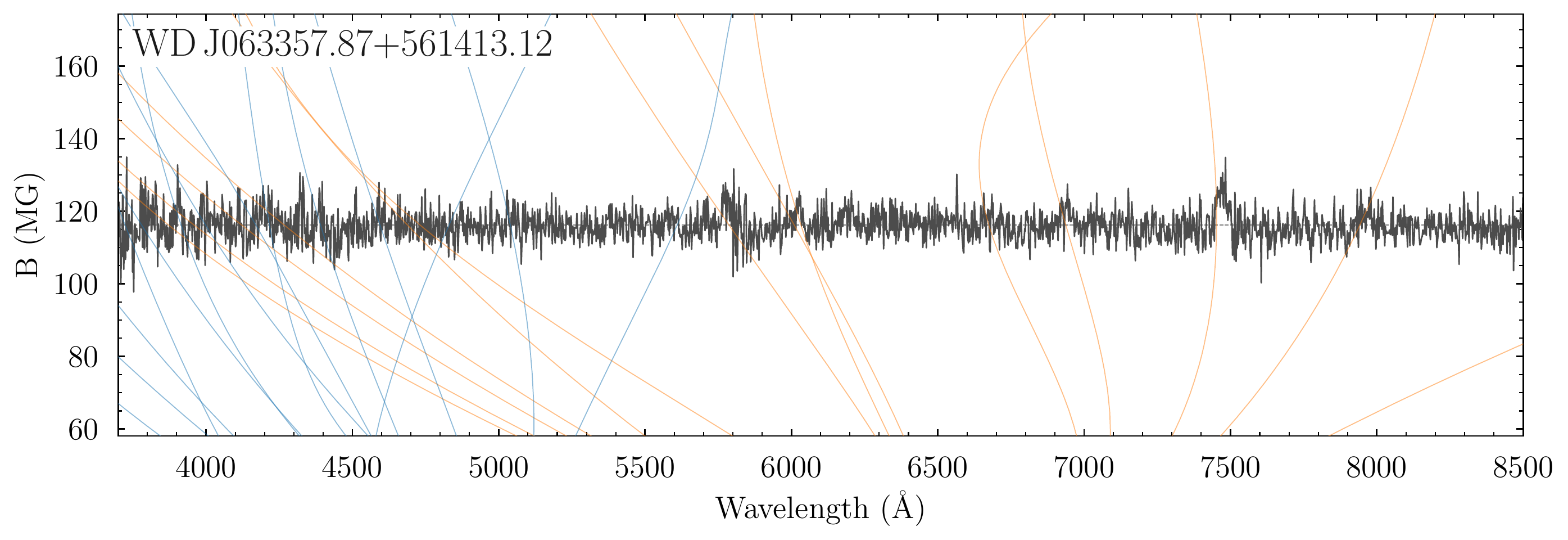}}
\centerline{\includegraphics[width=1.8\columnwidth]{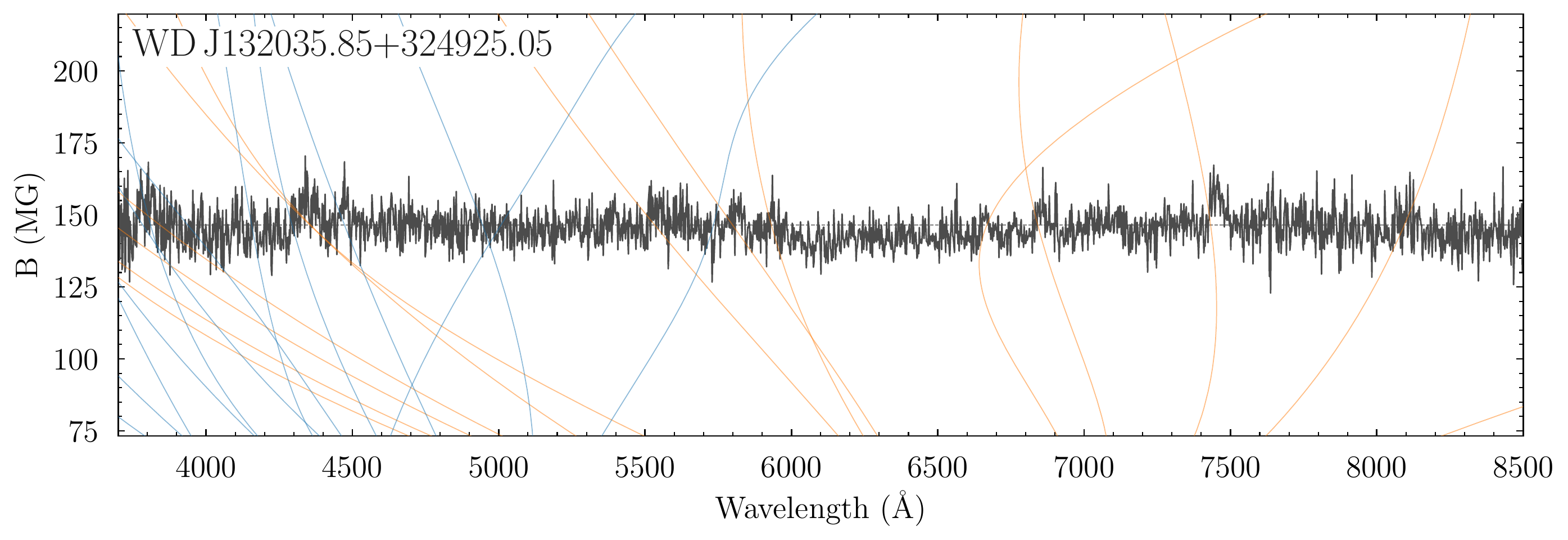}}
\caption{\label{f-normplot_appendix5} Continued.}
\end{figure*}




\bsp	
\label{lastpage}
\end{document}